%% file: SMP-20-015_temp.tex
\pdfoutput=1
\documentclass[11pt,twoside,a4paper,cmspaper,final,collab]{cms-tdr}
\def\svnVersion{df4b9d9}\def\svnDate{2022/04/18}\def\cmsCernNoTag{CERN-EP-2021-224}\def\cmsCernDate{\today}\def\cmsMessage{Published in Physical Review D as \href{http://dx.doi.org/10.1103/PhysRevD.105.092014}{\doi{10.1103/PhysRevD.105.092014}.}}
\begin{document}\cmsNoteHeader{SMP-20-015}

\newcommand{\MGfive} {\textsc{mg5}\_a\textsc{mc}\xspace}
\newcommand{\zcj} {\ensuremath{\PZ + \PQc\,\text{jet}}\xspace}
\newcommand{\zbj} {\ensuremath{\PZ + \PQb\,\text{jet}}\xspace}
\newcommand{\zlj} {\ensuremath{\PZ + \text{light jet}}\xspace}
\newcommand{\zljforbrac} {\ensuremath{\PZ + \text{light jet}}\xspace}
\newcommand{\zcjforbrac}{\ensuremath{\PZ + \PQc\,\text{jet}}\xspace}
\newcommand{\zbjs} {\ensuremath{\PZ + \PQb\,\text{jets}} \xspace}
\newcommand{\znjs} {\ensuremath{\Pp\Pp \to \PZ + n\,\text{jets}}\xspace}
\newcommand{\zjs}{\ensuremath{\PZ + \text{jets}}\xspace}
\newcommand{\zj}{\ensuremath{\PZ + \text{jet}}\xspace}
\newcommand{\wjs}{\ensuremath{\PW + \text{jets}}\xspace}
\newcommand{\ZpT}{\ensuremath{p_\mathrm{T}^{\PZ}}\xspace}
\newcommand{\bjetpT} {\ensuremath{p_\mathrm{T}^{\PQb\,\text{jet}}}\xspace}
\newcommand{\bjetabseta} {\ensuremath{\abs{\eta^{\PQb\,\text{jet}}}}\xspace} 
\newcommand{\deltaPhi} {\ensuremath{\Delta \phi^{(\PZ,\PQb\, \text{jet})}}\xspace}
\newcommand{\deltaY} {\ensuremath{\Delta y^{(\PZ,\PQb\,\text{jet})}}\xspace}
\newcommand{\deltaR} {\ensuremath{\Delta R^{(\PZ,\PQb\,\text{jet})}}\xspace}
\newcommand{\bbbarnospace}{\PQb{}\PAQb}
\newcommand{\zoneb}{\ensuremath{\PZ + 1\,\PQb\,\text{jet}}\xspace}
\newcommand{\ztwobs}{\ensuremath{\PZ + 2\,\PQb\, \text{jets}}\xspace}
\newcommand{\zgtoneb}{\ensuremath{\PZ + {\geq}\, 1\,\PQb\,\text{jet}}\xspace}
\newcommand{\zgttwob}{\ensuremath{\PZ + {\geq}\, 2\,\PQb\,\text{jets}}\xspace}
\newcommand{\emuonebj}{\ensuremath{\Pe\PGm + {\geq}\,1\,\PQb\,\text{jet}}\xspace}
\newcommand{\emutwobjs}{\ensuremath{\Pe\PGm + {\geq}\,2\,\PQb\,\text{jets}}\xspace}
\newcommand{\bj}{\PQb jet\xspace}
\newcommand{\zllgtonebj}{\ensuremath{\PZ (\to\Pell\Pell) + {\geq}\,1\,\PQb\,\text{jet}}\xspace}
\newcommand{\zllgttwobj}{\ensuremath{\PZ (\to\Pell\Pell) + {\geq}\,2\,\PQb\,\text{jets}}\xspace}
\newcommand{\Irel}{\ensuremath{\mathrm{I}_{\text{rel}}}\xspace}
\newcommand{\pTPU}{\ensuremath{p_{\mathrm{T}}^{\mathrm{PU}}}\xspace}
\providecommand{\cmsTable}[1]{\resizebox{\textwidth}{!}{#1}}
\providecommand{\cmsLeft}{left\xspace}
\providecommand{\cmsRight}{right\xspace}
\ifthenelse{\boolean{cms@external}}{\renewcommand{\NA}{\ensuremath{\cdots}}}{}

\cmsNoteHeader{SMP-20-015}

\title{Measurement of the production cross section for \texorpdfstring{\zbjs}{Z + b jets} in proton-proton collisions at \texorpdfstring{$\sqrt{s}=13\TeV$}{sqrt{s} = 13 TeV}}

\date{\today}

\abstract{
The measurement of the cross section for the production of a \PZ boson, decaying to dielectrons or dimuons, in association with at least one bottom quark jet are performed with proton-proton collision data at $\sqrt{s}=$ 13\TeV. The data sample corresponds to an integrated luminosity of 137\fbinv, collected by the CMS experiment at the LHC during 2016--2018. The integrated cross sections for \zgtoneb and \zgttwob are reported for the electron, muon, and combined channels. The fiducial cross sections in the combined channel are $6.52\pm 0.04\stat\pm 0.40\syst\pm 0.14\thy$\unit{pb} for \zgtoneb and $0.65\pm 0.03\stat\pm 0.07\syst\pm 0.02\thy$\unit{pb} for \zgttwob. The differential cross section distributions are measured as functions of various kinematic observables that are useful for precision tests of perturbative quantum chromodynamics predictions. The ratios of integrated and differential cross sections for \zgttwob and \zgtoneb processes are also determined. The value of the integrated cross section ratio measured in the combined channel is $0.100\pm 0.005\stat\pm 0.007\syst\pm 0.003 \thy$. All measurements are compared with predictions from various event generators.
}

\hypersetup{
pdfauthor={CMS Collaboration},
pdftitle={Measurement of the production cross section for Z + b jets in proton-proton collisions at sqrt(s) = 13 TeV},
pdfsubject={CMS},
pdfkeywords={CMS, Z boson, jets}}

\maketitle

\section{Introduction}
\begin{figure*}[htb!]
\centering
{\includegraphics[width=0.65\textwidth]{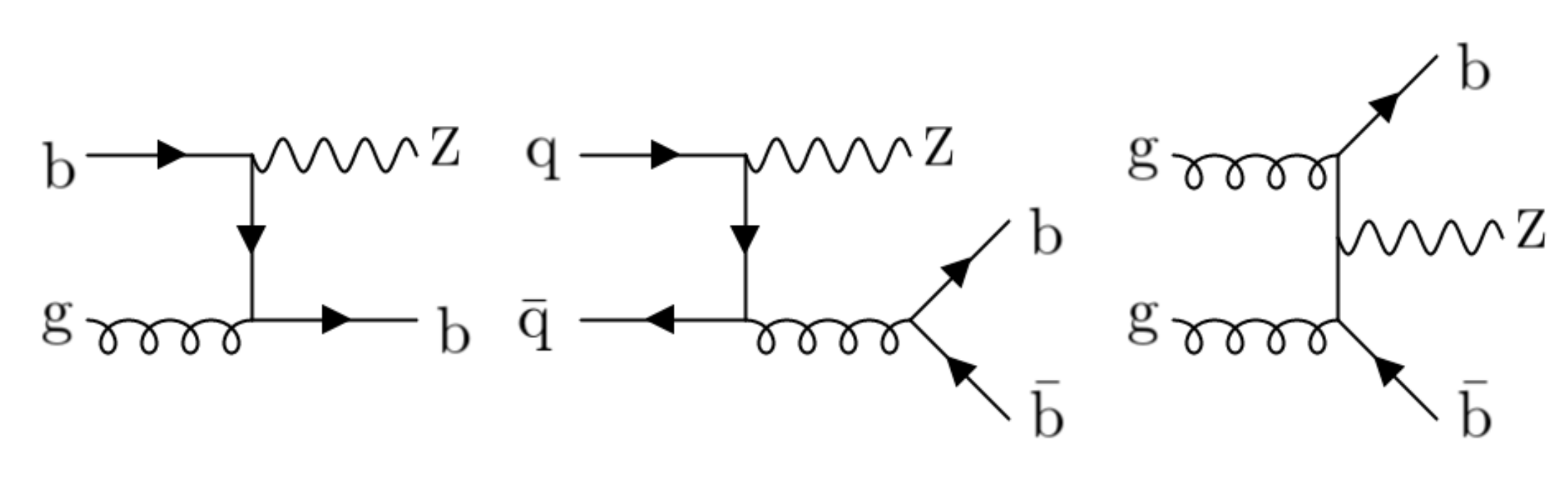}}
\caption{Examples of Feynman diagrams for \zoneb (left) and \ztwobs (middle and right).}
\label{fig:Z+2bjets}
\end{figure*}

A \PZ boson accompanied by bottom (\PQb) jets can originate from quark-gluon, quark-antiquark, and gluon-gluon interactions in proton-proton (pp) collisions. These processes can be used to test perturbative quantum chromodynamics (pQCD) predictions by comparing the measured cross sections with theoretical predictions. They also provide information on the \PQb quark parton distribution functions (PDFs)~\cite{HFcontent}. The measurement of \zbjs helps to estimate the uncertainty coming from the PDF choice in the {\PW} boson mass measurement~\cite{PhysRevD.83.113008WMass}. These processes are the dominant background for Higgs boson production in association with a \PZ boson ($\PZ\PH$, \PH $\to$ \bbbarnospace) in the standard model. They are also a source of background in many scenarios beyond the standard model such as the production of supersymmetric Higgs bosons in association with a \PQb quark, and new generations of heavy quarks decaying to a \PZ boson and a \PQb quark. Examples of Feynman diagrams of the signal process with quark-gluon, quark-antiquark, and gluon-gluon interactions for \zoneb and \ztwobs final states are shown in Fig.~\ref{fig:Z+2bjets}. Two different approaches are available for calculating the \zbjs production, the five-flavor and four-flavor schemes, and, as shown in Ref.~\cite{Maltoni}, they both yield consistent results.

Previous measurements have been performed on \zbjs at $\sqrt{s}= 1.96\TeV$ by the CDF~\cite{CDF1.96ppbar} and \DZERO~\cite{D01.96ppbar} Collaborations at the Fermilab Tevatron in proton-antiproton collisions. At the CERN LHC similar measurements have been performed at 7\TeV by the CMS~\cite{JHEP10(2014)141Zb7TeVCMS} and ATLAS~\cite{JHEP10(2014)141Zb7TeVATLAS} Collaborations, at 8\TeV by CMS~\cite{Zb8TeVCMS}, and at 13\TeV by ATLAS~\cite{Zb13TeVATLAS}, all in pp collisions.

In this paper, the integrated cross sections together with differential distributions of the \zllgtonebj and \zllgttwobj are reported (where $\Pell\Pell$ = $\Pe\Pe$ or $\PGm\PGm$). The differential distributions normalized to the corresponding fiducial cross sections are presented as well. In the measurements of normalized distributions, several systematic uncertainties cancel, which allows a more precise comparison with theory. The differential and normalized differential cross section distributions for \zgtoneb production are measured with respect to six kinematic observables: the transverse momentum of the dilepton (\ZpT); the highest \pt (leading) \PQb jet (\bjetpT); the absolute pseudorapidity of the \PQb jet (\bjetabseta); the azimuthal angle difference between the \PZ boson and the \PQb jet (\deltaPhi); the rapidity difference between the \PZ boson and the \PQb jet (\deltaY); and the angular correlation variable, which is defined as the separation in the pseudorapidity and azimuthal distance between the \PZ boson and the \PQb jet $[\Delta R^{(\PZ,\PQb\,\text{jet})}=\sqrt{\smash[b]{\bigl(\Delta \eta^{(\PZ,\PQb\,\text{jet})}\bigr)}^2 + \smash[b]{\bigl(\Delta \phi^{(\PZ,\PQb\,\text{jet})}\bigr)^2}}]$.

The differential and normalized differential cross sections for the \zgttwob production are measured as functions of nine kinematic observables describing the final state: the transverse momenta of the \PZ boson and the two highest \pt (leading and subleading) \PQb jets; the absolute pseudorapidity of the highest \pt \PQb jet; the angular correlations between the \PZ boson and the \PQb jets; the invariant mass of the two \PQb jets ($m_{\PQb\PQb}$); the invariant mass of the \PZ boson and the two \PQb jets ($m_{\PZ\PQb\PQb}$); the angular separation between two \PQb jets (${\Delta R_{\PQb\PQb}}$); the minimum separation between the \PZ boson and the two \PQb jets ($\Delta R^{\text{min}}_{\PZ\PQb\PQb}=\text{min}(\Delta R_{\PZ,\PQb_1},\Delta R_{\PZ,\PQb_2})$, where $\PQb_1$ and $\PQb_2$ are the two leading \PQb jets); and the asymmetry of the \zgttwob system ($A_{\PZ\PQb\PQb}=[\Delta R^\text{max}_{\PZ\PQb\PQb}-\Delta R^\text{min}_{\PZ\PQb\PQb}]/[\Delta R^\text{max}_{\PZ\PQb\PQb}+\Delta R^\text{min}_{\PZ\PQb\PQb}]$, where $\Delta R^\text{max}_{\PZ\PQb\PQb}=\text{max}(\Delta R_{\PZ,\PQb_1},\Delta R_{\PZ,\PQb_2})$ is the maximum separation between the \PZ and the two \PQb jets).

In addition, the integrated cross section ratios of \zgttwob to \zgtoneb are measured. The differential cross section ratios are measured as functions of \pt and $\abs{\eta}$ of the leading \PQb jet. These measurements allow a further reduction in various experimental uncertainties such as those related to the jet energy scale measurement. 

This study at higher $\sqrt{s}$ allows the exploration of the more extreme regions of phase space. The measurement of \bjetpT and \ZpT are sensitive to the \PQb quark PDF and the details of the pQCD calculations. The angular separation is sensitive to initial- and final-state radiation and multi-parton interactions where extra jet(s) affect the corresponding variables~\cite{Theor_ISR_FSR_MPI}. The invariant mass distributions are sensitive to two-\PQb-jet resonances. Measurements are compared with the predictions from various Monte Carlo (MC) event generators to test the higher-order pQCD calculations and simulations of parton showering, hadronization, and multiparton interactions.

Tabulated results are provided in the HEPData record for this analysis~\cite{hepdata}. 

The outline of this paper is as follows. Section 2 gives a brief description of the CMS detector. Section 3 describes the data and simulated samples used for the validation and unfolding studies, and Section 4 describes the event selection criteria. The unfolding procedure and systematic uncertainties are discussed in Section 5, and the results are presented in Section 6. Finally, the paper is summarized in Section 7.

\section{The CMS detector}
The central feature of the CMS apparatus is a superconducting solenoid of 6\unit{m} internal diameter, providing a magnetic field of 3.8\unit{T}. Within the solenoid volume there is a silicon pixel and strip tracker covering a pseudorapidity region $\abs{\eta} < 2.5$ (3) in 2016 (2017--2018) data-taking period~\cite{trackerup}, a lead tungstate crystal electromagnetic calorimeter (ECAL), and a brass and scintillator hadron calorimeter. Forward calorimeters, made of steel and quartz fibers, extend the coverage to $\abs{\eta} < 5.0$ provided by the barrel and endcap detectors. Muons are detected in gas-ionization chambers embedded in the steel flux-return yoke outside the solenoid. Events of interest are selected using a two-tiered trigger system~\cite{Khachatryan:2016bia}. The first level (L1), composed of custom hardware processors, uses information from the calorimeters and muon detectors to select events at a rate of around 100\unit{kHz} within a latency of less than 4\unit{$\mu$s}~\cite{L1trigger}. The second level, known as the high-level trigger, consists of a farm of processors running a version of the full event reconstruction software optimized for fast processing and reduces the event rate to around 1\unit{kHz} before data storage.

A more detailed description of the CMS detector, together with a definition of the coordinate system used and the relevant kinematic variables, is reported in Ref.~\cite{Chatrchyan:2008zzk}.

\section{Data and simulated samples}\label{sec:samples}
This analysis is performed using a data sample of $\Pp\Pp$ collisions at a center-of-mass energy of 13\TeV collected by the CMS experiment during the 2016 (35.9\fbinv), 2017 (41.5\fbinv), and 2018 (59.7\fbinv) data-taking periods with an integrated luminosity of 137\fbinv. 

Several Monte Carlo event generators are used to simulate signal and background processes. The Drell--Yan (DY) process with exclusive jet multiplicity up to 2 is simulated at next-to-leading order (NLO) precision by \MGvATNLO (denoted \MGfive)~\cite{Alwall:2014hca} version 2.3.2.2 for 2016 data and version 2.6.0 for the 2017--2018 data with the FXFX~\cite{FXFX} matching between the jets from matrix element calculations and parton showers. The NNPDF 3.0 NLO and NNPDF 3.1 next-to-NLO (NNLO) PDF sets~\cite{pdfset} are used for the 2016 and 2017--2018 data-taking periods, respectively. 

A second sample is produced with \MGfive version 2.2.2 (2.4.2) for the 2016 (2017--2018) data-taking period using leading order (LO) matrix elements for \znjs ($n \leq 4$), with the MLM matching scheme~\cite{MLM}. The NNPDF 3.0 LO and NNPDF 3.1 NNLO PDF sets are used for the 2016 and 2017--2018 data-taking periods, respectively. The \MGfive generator is interfaced with {\PYTHIA} v8.212 (v8.230)~\cite{pythia8.2} for parton showering, hadronization, and the underlying event simulation using tune CUETP8M1~\cite{Khachatryan:2015pea} (CP5~\cite{CP5_tune}) for the 2016 (2017--2018) data-taking period. Matching between the matrix element generators and the parton shower jets is done with the matching scales set at 19 and 30\GeV for the LO and NLO \MGfive generators, respectively. 
 
A third inclusive sample has been produced with \SHERPA~v2.2.4~\cite{Bothmann:2019yzt} to generate \znjs events, with $n \leq 2$ at NLO and $n = 3, 4$ at LO. The merging with the \SHERPA parton shower is done via the MEPS@NLO prescription~\cite{Hoeche:2012yf,Gehrmann:2013,Hoeche:2014rya} with a matching scale of 20\GeV. The NNPDF 3.0 NLO PDF and a dedicated set of tuned parton shower parameters developed by the \SHERPA authors are used. In the matrix element calculation, the value of the NNPDF 3.0 strong coupling (\alpS) at the \PZ boson mass is set to 0.130 (0.118) for \MGfive at LO and is set to 0.118 (0.118) for \MGfive at NLO for the 2016 (2017--2018) data-taking period. All three signal samples use the five-flavor number scheme with \PQb quark density allowed in the initial state via a b and charm (\PQc) quark PDF, with the \PQb and \PQc quarks typically being massless.

The DY process includes signal events with a \PZ boson and a jet initiated by a \PQb quark (\PQb jet) as well as background events from \zcj and \zlj in which a jet is initiated by a charm quark (\PQc jet) and a light quark or gluon (light jet), respectively. All DY processes are scaled to NNLO accuracy in QCD based on the \zjs cross sections calculated with \FEWZ v3.1~\cite{Li_2012}. Other background contributions come from top quark-antiquark (\ttbar) and single top quark ($s$- and $t$-channel, and $\cPqt\PW$) production as well as from diboson ($\PW\PW$, $\PW\PZ$, and $\PZ\PZ$) processes. The \ttbar process is generated at NLO by \POWHEG v2.0~\cite{powheg1,powheg2,powheg3} in 2016 and \MGfive in 2017--2018 data-taking periods and interfaced with {\PYTHIA} v8.212 (CUETP8M2T4~\cite{CMS:2016kle}), v8.226 (CP5), and v8.230 (CP5) for 2016, 2017, and 2018, respectively, for parton showering and hadronization. The single top quark $t$-channel and $\cPqt\PW$ processes are simulated at NLO with \POWHEG v2.0 and \POWHEG v1.0, respectively, whereas the $s$-channel process is simulated with \MGfive at NLO, both interfaced with {\PYTHIA}. The {\PYTHIA} v8.212, v8.212, and v8.205 are used for single top quark $t$-channel, $\cPqt\PW$, and $s$-channel processes, respectively, in 2016, while {\PYTHIA} v8.226 and v8.230 are used in 2017 and 2018, respectively, for all single top quark process. The diboson background processes are simulated at LO using {\PYTHIA} v8.205, v8.226, and v8.230 for 2016, 2017, and 2018, respectively. 

All simulation samples, except those based on \SHERPA, use {\PYTHIA} with the CUETP8M1 or CUETP8M2T4 or CP5 tune. The CUETP8M1 tune includes the LO NNPDF 2.3~\cite{Ball_2013} PDFs set with the strong coupling $\alpS (m_{\PZ})$ set 0.137 for space- and time-like shower simulation. The CUETP8M2T4 tune is based on the CUETP8M1 tune, which includes the NNPDF30\_lo\_as\_0130 PDF set, but uses a lower value of $\alpS = 0.111$ for the initial-state radiation component of the parton shower. The CP5 tune use the NNPDF3.1 PDF set at NNLO, with $\alpS$ values of 0.118, and running according to NLO evolution. The diboson and single top quark predictions are scaled to NNLO accuracy~\cite{NNLO_computation,Kidonakis:2013zqa}. The prediction for \ttbar production is normalized to NNLO calculations in pQCD including resummation of next-to-next-to-leading logarithmic soft-gluon terms with \TOPpp 2.0~\cite{Czakon:2013goa,Baernreuther:2012ws,Beneke:2011mq,Cacciari:2011hy,Czakon:2012zr,Czakon:2012pz,Czakon:2011xx}.

The MC samples are processed with the full CMS detector simulation based on \GEANTfour~\cite{Agostinelli:2002hh}. The simulated events are reconstructed using the same algorithms used for the data, and include additional interactions per bunch crossings, referred to as pileup (PU). Simulated events are weighted so that the PU distribution reproduces the one observed in data, which has an average of about 23 (32) interactions per bunch crossing in 2016 (2017--2018).

\section{Event selection}\label{sec:objects}
The final state in this analysis contains a \PZ boson ($\PZ \to \Pe\Pe/\PGm\PGm$) and at least one jet coming from a \PQb quark. Therefore, the measurement requires the reconstruction of electrons, muons, and \PQb jets. Events are reconstructed using a particle-flow (PF) algorithm~\cite{pf}, which identifies each individual particle candidate using information from various subdetectors. Energy deposits are measured in the calorimeters, and charged particles are identified in the central tracking and muon systems.

The collision events are required to pass single lepton triggers. They must have at least one electron (muon) candidate with a minimum $\pt$ of 27, 32, and 32\GeV (24, 27, and 24\GeV) during the 2016, 2017, and 2018 data-taking periods, respectively.

The candidate vertex with the largest value of summed physics-object $\pt^2$ is the primary $\Pp\Pp$ interaction vertex. The physics objects are the leptons and jets, clustered using the anti-\kt algorithm~\cite{Cacciari:2008gp} with the tracks assigned to candidate vertices as inputs, and the associated missing transverse momentum (\ptvecmiss), the negative vector \pt sum of all tracks. The leptons in the analysis are required to originate from the primary vertex.

Electrons are identified as a primary charged particle track and potentially many ECAL energy clusters. These clusters correspond to the electron track extrapolation to the ECAL and to possible bremsstrahlung photons emitted by electrons along the way through the tracker material~\cite{Khachatryan:2015hwa}. The electron momentum is estimated by combining the energy measurement in the ECAL with the momentum measurement in the tracker. The longitudinal (transverse) impact parameters with respect to the beam line for the barrel ($\abs{\eta} < 1.48$) and endcap ($1.48 < \abs{\eta} < 3.0$) regions are required to be less than 0.10 (0.05)\unit{cm} and 0.20 (0.10)\unit{cm}, respectively. The momentum resolution for electrons with $\pt\approx45\GeV$ ranges from 1.7 to 4.5\% depending on $\eta$. It is generally better in the barrel region than in the endcaps, and also depends on the bremsstrahlung energy emitted by the electron as it traverses the material in front of the ECAL. The dielectron mass resolution for $\PZ\to\Pe\Pe$ decays when both electrons are in the ECAL barrel is 1.9\%, and degrades to 2.9\% when both electrons are in the endcaps.

Muons are measured in the pseudorapidity range $\abs{\eta} < 2.4$. The single muon trigger efficiency exceeds 90\% over the full $\eta$ range, and the efficiency to reconstruct and identify muons is greater than 96\% for  $\pt > 25\GeV$. Efficiencies are measured using the data-driven tag-and-probe methods. Matching muons to tracks measured in the silicon tracker results in a transverse momentum resolution for muons with \pt up to 100\GeV of 1\% in the barrel and 3\% in the endcaps~\cite{Sirunyan:2018}. The corrections for the muon transverse momentum scale and resolution~\cite{roch} are also applied to remove biases of the muon momentum from detector misalignments or magnetic field uncertainties. 

To reduce the misidentification rate, leptons are required to be isolated. The relative isolation variable is defined as:
\begin{linenomath}
\begin{equation}
\Irel = \frac{\sum p_{\mathrm{T}i}^{\text{charged}} + \max(0, \sum E_{\mathrm{T}i}^{\text{neutral}} + \sum E_{\mathrm{T}i}^{\gamma}- \pTPU)}{\pt^{\ell}} , 
\end{equation}
\end{linenomath}
which is the sum of the \pt values for all PF candidates within a cone of radius $\Delta R =$ 0.3 (0.4) centered on the electron (muon) track direction. Here, $\sum E_{\mathrm{T}i}^{\smash[b]{\, \text{neutral}}}$ and $\sum E_{\mathrm{T}i}^{\gamma}$ are the scalar sums of the transverse energies of neutral hadrons and photons, respectively. The quantity $\sum p_{\mathrm{T}i}^{\smash[b]{\, \text{charged}}}$ represents the \pt sum of the charged hadrons associated with the primary vertex and \pTPU is the contribution from PU. For electrons, \pTPU is evaluated using the ``jet area'' method described in~\cite{Cacciari_2008_PU_sub}. For muons, \pTPU is taken to be half of the scalar \pt sum deposited in the isolation cone by charged particles not associated with the primary vertex. The factor of one half corresponds approximately to the ratio of neutral to charged hadrons produced in the hadronization of PU interactions. Finally, $\pt^{\ell}$ stands for the lepton (electron or muon) transverse momentum. For muons \Irel is required to be less than 0.15. For electrons \Irel is \pt dependent; for 25\GeV $\Irel < 0.021\,(0.040)$ and for 35\GeV $\Irel < 0.015\,(0.029)$ in the barrel (endcap) region~\cite{e_recons_2021}.

{\tolerance=800 The small residual differences in the lepton (electron and muon) trigger, identification, and isolation efficiencies between data and simulation are measured using a ``tag-and-probe'' method~\cite{tagprob} and are included by applying scale factors to simulated events.\par}
 
Jets are clustered from PF candidates using the infrared- and collinear-safe anti-\kt algorithm with a distance parameter of 0.4, as implemented in the \FASTJET package~\cite{Cacciari_2012}. The jet momentum is the vector sum of all particle momenta in the jet, and is found from simulation to be within 5--10\% of the true momentum over the entire \pt spectrum and detector acceptance. Pileup interactions could result in additional tracks and calorimetric energy depositions, possibly increasing the measured jet momentum. To mitigate this effect, tracks identified as originating from PU vertices are discarded and an offset correction~\cite{Cacciari_2008_PU_sub} is applied to correct for remaining contributions. 

The reconstructed jet energy scale (JES) is corrected using a factorized model to compensate for the nonlinear and nonuniform response in the calorimeters. Corrections are derived from simulation to bring, on average, the measured response of jets to that of particle-level jets. In situ measurements of the momentum balance in dijet, multijet, photon $+$ jet, and leptonically decaying \zj events are used to account for any residual differences between the JES in data and simulation~\cite{cmsJEC}. The jet energy resolution (JER) in the simulation is smeared to reproduce what is observed in data. The JER amounts typically to 16\% at 30\GeV and 8\% at 100\GeV. Additional selection criteria are applied to remove jets potentially dominated by anomalous contributions from various subdetector components or reconstruction failures~\cite{CMS-PAS-JME-16-003}. Jets identified as likely to originate from PU are also removed by using a pileup jet ID discriminator for jets with $30 < \pt \leq 50\GeV$~\cite{Sirunyan_2020}. For jets with $\pt > 50\GeV$, the pileup jet ID discriminator is not applied, since here the pileup event contribution is negligible~\cite{CMS-PAS-JME-13-005}.

The missing transverse momentum vector \ptvecmiss is computed as the negative vector sum of the transverse momenta of all the PF candidates in an event, and its magnitude is denoted as \ptmiss~\cite{Sirunyan:2019kia}. The \ptvecmiss is modified to include corrections to the energy scale of the reconstructed jets in the event.

A neural network algorithm (\textsc{DeepCSV})~\cite{deepCSV} is used to discriminate \PQb jets from \PQc and light jets. In the \textsc{DeepCSV} algorithm, the significance of the secondary vertex displacement is combined with track-based information in a machine-learned multivariate analysis to increase the efficiency compared with cut-based algorithms.

The \PZ boson candidate is reconstructed from two same flavor leptons with opposite charge having invariant mass ($m_{\ell\ell}$) between 71--111\GeV. Here, leading and subleading leptons must have $\pt > 35\GeV$ and $\pt > 25\GeV$, respectively. The \PZ boson event must contain at least one \PQb jet with $\pt > 30\GeV$, $\abs{\eta} < 2.4$, and have $\ptmiss < 50\GeV$. The \PQb jets are selected with the tight operating point of the \PQb tag discriminator. The overlap between the \PQb jet and a lepton from the \PZ boson decay is removed by requiring a minimum distance of $\Delta R > 0.4$ between them.

In simulation, the classification of reconstructed \zj events into \zbj, \zcj, and \zlj categories is based on the flavors of reconstructed jets with $\pt > 30\GeV$ and $\abs{\eta} < 2.4$. They are classified as \PQb or \PQc jets if they are matched to the generated \PQb or \PQc hadrons. The matching is done using the ghost association procedure in which the modulus of the hadron four-momentum is set to a small number, retaining only the directional information~\cite{deepCSV}. In the case when both \PQc and \PQb hadrons are matched, the jet is considered as a \PQb jet. Based on reconstructed jets with defined flavors, events are classified as \zbj events if they contain at least one \bj. Of the remaining events, those that contain at least one \PQc hadron are considered as \zcj events and those that contain neither \PQb nor \PQc hadrons are classified as \zlj events. In this analysis the tagging efficiency of \zbj (\zcj) events is estimated as the fraction of events that pass the \PQb (\PQc) hadron and \PQb tagging (\PQc tagging) requirement over the number of events that pass just the \PQb (\PQc) hadron requirement. In simulations for the tight operating point, the tagging efficiencies for \PQb (\PQc) quarks are 50--60\% (25\%) and misidentification probabilities are about 0.1\% ($<$1.5\%) and 2--5\% ($\approx$24\%) for a light (light) and \PQc (\PQb) jet. 

Data from \ttbar, \wjs, and multijet events that are enriched with a given flavor of interest are used to measure the \PQb, \PQc, and light jet tagging efficiencies. These are compared with the efficiencies calculated from simulation. Small differences between data and simulation are corrected through scale factors applied to simulation.

The selected events with a \PZ boson and at least one \PQb jet have $\approx$18\% background contamination, with the major contributions arising from the \zcj ($\approx$7\%), \ttbar ($\approx$7\%), and \zlj ($\approx$4\%) processes. In the \zgttwob signal region a significant source of background originates from \ttbar production, accounting for $\approx$30\% of the events in the signal region and $\approx$85\% of the background. To validate the simulation, control regions that are enriched with processes of interests are constructed.

For the \zgtoneb measurement, the \ttbar control region is constructed by requiring an oppositely charged electron and muon pair with at least one \PQb jet (\emuonebj) with the same selection criteria discussed above. A sample containing about 92\% of \ttbar events is obtained for each year of data taking. The non-\ttbar contributions to this region are DY, single top, \wjs, and diboson events. A scale factor, (data $-$ non-\ttbar background)/(\ttbar MC), is extracted from this region. It is 0.98 for 2016 and 0.87 for 2017--2018 data-taking periods and is applied in the signal region to the \ttbar simulation in the $\Pe\Pe$ and $\PGm\PGm$ channel.

In the \zgttwob signal region, the \ttbar background is estimated by a data-driven method using an \emutwobjs sample where the purity of the \ttbar events is $\approx$96\%. A normalization factor is derived by comparing the numbers of \ttbar events seen in the sidebands of the dilepton invariant mass distributions from \emutwobjs and signal (\zgttwob) regions. This factor is applied to the number of \ttbar ($\to\Pe\PGm$) $+$ $\geq$ 2 \PQb jets events within the \PZ boson mass window to obtain the \ttbar background contribution in the signal region.

The \zlj control region (CR1) is constructed by requiring events to have inclusive jets without any condition on the tagger discriminator. A sample containing about 83--85\% of \zlj events is obtained for each year of data taking. The non-(\zljforbrac) backgrounds are \zcj, \zbj, diboson, \ttbar, single top, and \wjs events. A scale factor, (data $-$ non-(\zljforbrac) background)/(\zlj MC), is extracted from this region and applied in the signal region to the \zlj MC. The scale factors 0.94, 0.98, and 0.90 are used for the 2016, 2017, and 2018 data-taking periods, respectively. As a cross-check, another \zlj control region (CR2) is defined by requiring events to pass anti-\PQb and anti-\PQc tagging tight operating points of \PQb and \PQc quark discriminators. The scale factors in CR1 and CR2 are consistent within 1--2\%.

The \zcj control region (CR3) is constructed by requiring \PQc jet events with the tight operating points of the \PQc tag discriminator. A sample containing about 41--55\% of \zcj events is obtained for each year of data taking. Scale factors, (data $-$ non-(\zcjforbrac) background)/(\zcj MC), 0.75, 1.08, and 0.99 for 2016, 2017, and 2018 data-taking periods, are extracted from this region and are applied to the \zcj MC sample in the signal region. The non-\zcj contributions to this region are \zbj, \zlj, \ttbar, single top, \wjs, and diboson events. As a cross-check, another \zcj control region (CR4) is defined by requiring events to pass tight operating points of the \PQc quark discriminator and inverted \PQb quark discriminator. The scale factors in CR3 and CR4 are consistent within 1--2\%.

The collective contribution of single top, \wjs, and diboson processes is small ($<$2\%). Thus, data-driven estimates are not considered for these processes and their contributions are determined from simulation.

\section{Unfolding and systematic uncertainties}
The reconstructed distributions are unfolded to generator-level quantities using the \textsc{TUnfold} package (v17.5)~\cite{Schmitt_2012}, which is based on a least squares fit, to correct for the detector resolution and the selection efficiencies. The simulated events are classified according to generator-level information. The generator-level jets are formed from stable particles (c$\tau > 1\unit{cm}$), except neutrinos, using the same anti-\kt algorithm as for the reconstructed jets. Any overlap between generator-level jets and a pair of leptons from the \PZ boson decays is removed by requiring a minimum distance of 0.4 between them. If an event consists of at least one (two) generator-level jet(s) containing a \PQb hadron, the event is classified as a \zgtoneb (\zgttwob) event. The generator-level leptons are ``dressed'' by adding the momenta of all photons within $\Delta R \leq 0.1$ around the lepton direction to account for the final-state radiation effects. The distributions are unfolded to the generator-level in the fiducial region defined in Table~\ref{table:1}.

The reconstructed distributions are unfolded with a response matrix that describes the migration probability between the particle- and reconstruction-level quantities.
The response matrix is constructed by spatially matching the reconstruction-level leptons and \PQb jet objects to the corresponding generator-level objects within $\Delta R <$ 0.3. The selected \zgtoneb and \zgttwob samples at reconstruction-level can have background events that do not match to generator-level objects. The background events are channel dependent and they are subtracted from reconstructed events for each year and channel separately. The background subtracted reconstruction-level distributions are unfolded. The unfolded distributions are corrected for those events that have generator-level objects in the fiducial volume, which do not match with reconstructed objects. Events generated with \MGfive (NLO) and CP5 tune are used to construct the response matrix. The \MGfive (NLO) with CUETP8M1 tune, \MGfive (LO) with CP5 and CUETP8M1 tunes, and \SHERPA are used for comparison with unfolded data.
 
The effect of regularization, a procedure to control large fluctuations of unfolded distributions due to statistical uncertainties of observed quantities~\cite{Tikhonov:1963}, is negligible for the selected observables. Therefore, all studies are performed without regularization. The unfolding procedure is validated with statistically independent tests in which \MGfive NLO distributions are unfolded with LO \MGfive and compared with generator-level NLO \MGfive predictions. There is good agreement between the unfolded and the generator-level predictions. The resolution of variables to be measured in the present analysis is not expected to differ for the electron and muon channels. This procedure is validated by unfolding generator-level MC distributions for the electron channel with the response matrix constructed for the muon channel. The background and acceptance distributions are taken from the electron channel only. The resulting unfolded distributions are compared with the unfolded results when the electron channel is used for the construction of the response matrix. A good agreement is observed between the two unfolded results, which establishes that the response matrices for the two channels are consistent with each other. Therefore, the electron and muon data are added and unfolded with the combined response matrix.
 
The data distributions for the \zgtoneb (\zgttwob) analysis are unfolded with a response matrix constructed using the \zbj events in the DY sample that is simulated with the \MGfive (NLO) generator.
 
\begin{table*}[ht!]
\centering

\topcaption{Fiducial region definition at generator-level.}
\begin{scotch}{lc}
 Object & Selection\\
\hline
  Dressed leptons & $\pt \text{(leading)} > 35\GeV$, $\pt \text{(subleading)} > 25\GeV$, $\abs{\eta} < 2.4$ \\
 \PZ boson & $71 < m_{\ell\ell} <  111\GeV$\\
  Generator-level \PQb jet &  \PQb hadron jet, \pt $>$ 30\GeV, $\abs{\eta} < 2.4$ \\
\end{scotch}
\label{table:1}
\end{table*}

\subsection{Experimental uncertainties}
There are a number of sources of uncertainty that have a significant impact on the cross section measurements. These include
\begin{itemize}
\item \textit{Integrated luminosity}: The integrated luminosities of the 2016--2018 data-taking periods have uncertainties in the 1.2--2.5\% range. These uncertainties are partially correlated and correspond to a total uncertainty of 1.6\% for the Run 2 (2016--2018)~\cite{CMS-LUM-17-003,CMS-PAS-LUM-17-004,CMS-PAS-LUM-18-002}. The luminosity uncertainty affects the integrated cross section but not the normalized distributions.

\item \textit{Jet energy scale and resolution}: The effect of JES and JER is evaluated by varying up and down the \pt values of jets with the corresponding uncertainty factors. The JES and JER uncertainties are also propagated to \ptmiss. The JES and JER scale factors are correlated with the JES and JER components of \ptmiss, so scale factors are varied up and down simultaneously. For the \zgtoneb (\zgttwob) analysis, the JES introduces a systematic uncertainty of $\approx$3\% ($\approx$6\%) on the integrated cross section whereas the effect of JER is less than 1\%. The JES has varying effects on the differential and normalized differential cross section distributions for the different observables of \zgtoneb (\zgttwob) events, with maximum effects of 9 and 7\% (24 and 18\%), respectively, in the low event count bins.

\item \textit{\PQb tagging/mistagging}: The uncertainty due to \PQb tagging and mistagging scale factors is evaluated by varying them up and down within their uncertainties. The scale factor for tagging \PQb jets is correlated with the \PQc jets scale factor, so for tagging \PQb and mistagging \PQc jets, scale factors are varied up and down simultaneously, while the scale factor for mistagging light jets is varied independently. The combined \PQb tagging uncertainty is then calculated as the sum in quadrature of these variations. For the \zgtoneb (\zgttwob) analysis the uncertainties in the \PQb tagging and mistagging scale factors affect the cross section measurements by 3.0\% (5.8\%). The differential and normalized differential cross section distributions are affected by 2--8\% and 0.5--5.0\%, respectively, for selected observables of \zgtoneb events. These uncertainties are less than 14.5 and 9.5\% in the \zgttwob differential and normalized differential cross section measurements, respectively. 

\item \textit{Unclustered energy of \ptmiss}: The effect of the uncertainty in the remaining component of \ptmiss, that is unclustered energy, is estimated by varying its contributions within the uncertainties, which are treated as correlated across years and channels. For \zgtoneb (\zgttwob), this uncertainty source affects the integrated cross section by up to 2.8\% (3.6\%). The differential and normalized differential cross section distributions of different observables of \zgtoneb (\zgttwob) events are affected by up to 3.8\% (8.4\%) and 1.1\% (7.1\%), respectively.

\item \textit{Lepton selection}: Electrons and muons are corrected for trigger, reconstruction, and identification efficiencies and the effects of these corrections are estimated by varying the corresponding scale factors within their uncertainties. For the \zgtoneb (\zgttwob) measurement, the uncertainty in the electron selection efficiency results in an uncertainty in the integrated cross section of about 4.6\% (4.3\%) for the electron channel, but for the combined channel the effect is about 1.5\% (1.4\%). Normalized distributions are not biased due to electron selection efficiency because associated uncertainties cancel in the ratio. The integrated cross section has a systematic uncertainty due to muon selection of less than 1\%. 

\item \textit{Pileup reweighting}: The PU distribution in the simulated samples is reweighted to match that of the data. The corresponding uncertainty is estimated by varying the total inelastic cross section by $\pm 4.6\%$~\cite{lumi}. For \zgtoneb (\zgttwob) the uncertainty in the PU simulation affects the integrated cross section by 1.7--2.4\% (2.1--2.9\%). For the \zgtoneb (\zgttwob) differential and normalized differential cross section distributions the effects are less than 2.7\% (6.8\%) and 1.3\% (4.3\%), respectively.

\item \textit{Background estimation}: In the \zgtoneb cross section measurements the \zcj and \zlj contributions come from simulation and are validated in control regions as discussed in Section~\ref{sec:objects}. The difference between data and MC simulation is extracted from each control region and is applied as a scale factor for the \zcj and \zlj MC predictions in the signal region, respectively. These scale factors can have associated systematic uncertainties from experimental sources (JES, \PQb tagging, PU, lepton selection), although these are already considered while evaluating systematic uncertainties in the signal region. For all of the control regions the data and MC simulation values are within 10\%. Hence, an additional systematic uncertainty of 10\% is assigned to the background modeling. For \zgtoneb, the effect of these uncertainties is $\approx$2.2\% on the integrated cross section, whereas the effect of uncertainties related to \zcj and \zlj backgrounds is about 1.5\%. The corresponding effects on the differential and normalized differential cross section distributions range from 1.5 to 4.8\% and 0.5 to 3.3\%, respectively. In the \zgttwob measurements the background uncertainty originates from the \ttbar, Drell--Yan, diboson, and single top estimations. The \ttbar uncertainty is the dominant source, and is derived by comparing the normalization coefficients obtained using fits in the dilepton invariant mass and missing transverse momentum distributions. The background uncertainty in the \zgttwob integrated cross section is 2.4\%. In the \zgttwob differential and normalized differential cross sections this uncertainty is less than 4.9 and 2.7\%, respectively. 

\item \textit{Model dependence}: The unfolding model uncertainty is estimated by reweighting the signal MC (\MGfive at NLO) with a scale factor that is calculated by fitting the ratio of background-subtracted data to signal simulation of normalized distributions and using it as an alternative model for the unfolding. It affects the differential and normalized differential cross section distributions of different observables in the \zgtoneb analysis up to 1.5 and 1.2\%, respectively.

\item \textit{Pileup jet identification}: The PU jet identification criteria have different performance in the data and simulated events. This difference is corrected as a scale factor. For \zgtoneb and \zgttwob measurements, the effect of uncertainties associated with the scale factor is less than 0.7\% in the integrated cross section. The differential and normalized differential cross section distributions of different observables of \zgtoneb (\zgttwob) events are affected up to 0.5\% (1.2\%).

\item \textit{L1 prefiring}: During the 2016--2017 data-taking period there was a timing issue in the $2.0 < \abs{\eta} < 3.0$ region of the ECAL where L1 trigger primitives were mistakenly associated with the previous bunch crossing. This led to a self-vetoing of events with significant energy in the $2.0 < \abs{\eta} < 3.0$ region and a loss of trigger efficiency. Correction factors are determined from data for the efficiency loss and applied to the simulation of the dielectron and dimuon channels. The uncertainty due to the prefiring scale factors is evaluated by varying them up and down within their uncertainties. For the \zgtoneb and \zgttwob differential and normalized differential cross section distributions as well as on the integrated cross section the effects are less than 0.5\%.
\end{itemize}

In the cross section ratio measurement, the luminosity uncertainty is effectively cancelled. All other experimental uncertainties except the background estimation are evaluated as follows. For an uncertainty of interest, the \zgttwob and \zgtoneb cross sections are obtained for up and down systematic variations described above and the corresponding ratios are obtained. The uncertainty is derived by comparing these ratios with the nominal ones. The background estimations in the \zgttwob and \zgtoneb events are based on independent methods and control regions. Therefore, the background uncertainties are uncorrelated and derived from quadrature sums of those from the \zgttwob and \zgtoneb measurements.

\subsection{Theoretical uncertainties}
 The uncertainties coming from the theory calculations and MC generators are estimated as follows.
\begin{itemize}
\item \textit{Renormalization and factorization scales}: The scale uncertainties are estimated using a set of weights provided by the generator that corresponds to variations of renormalization ($\mu_{\mathrm{R}}$) and factorization ($\mu_{\mathrm{F}}$) scales by factors of 0.5 and 2. The unfolded distributions are obtained for all combinations, except for $\mu_{\mathrm{F}}$/$\mu_{\mathrm{R}}=$ 0.25 or 4, and their envelope is quoted as the uncertainty. The integrated cross sections for \zgtoneb (\zgttwob) have scale uncertainties of 2.6\% (2.5\%), 2.9\% (2.3\%), and 2.1\% (2.5\%) in the dielectron, dimuon, and combined channels, respectively. The uncertainties in the \zgtoneb (\zgttwob) differential distributions vary from 1 to 10\% (0.5 to 4.9\%) and for the normalized differential distributions they vary from 0.5 to 9.4\% (0.5 to 5.0\%). Similarly, uncertainties are evaluated for \MGfive (NLO) with CP5 tune generator-level distributions and the effect on the \zgtoneb (\zgttwob) integrated cross section at the generator-level is around 6.6\% (9.0\%). The renormalization and factorization scale uncertainties in the cross section ratios are the quadrature sum of those from the \zgttwob and \zgtoneb measurements.

\item \textit{PDF}: The uncertainty in the parton distribution functions is estimated using different Hessian eigenvectors of the NNPDF 3.1 PDF sets. Applying the master formula of the Hessian PDF error calculation~\cite{hessian}, this uncertainty is estimated on a bin-by-bin basis for the differential distributions. The integrated cross sections for the \zgtoneb (\zgttwob) process have PDF uncertainties around 0.3--0.4\% (0.3\%) in the dielectron, dimuon, and combined channels. For the \zgtoneb (\zgttwob) analysis, the uncertainties in the differential distributions for the dielectron, dimuon, and combined channels are 0.5 to 4.2\% (0.5--4.0\%), whereas the uncertainties in the normalized differential cross sections for these channels are in the range 0.5--3.3\% (0.5--3.2\%). The PDF uncertainties are treated as correlated across the years and channels. Similarly, the effect of PDF (NNPDF 3.1) uncertainty in the generator-level distributions is obtained on a bin-by-bin basis of generator-level distributions with PDF weights applied. The effect on the integrated cross sections of the \zgtoneb and \zgttwob at the generator-level is around 1.0\%. The \zgttwob and \zgtoneb PDF uncertainties are summed in quadrature for the cross section ratios. The uncertainties in the PDFs are also estimated for \SHERPA at generator-level using the 102 replicas of the NNPDF 3.0 PDF set. The integrated cross sections for the \zgtoneb (\zgttwob) process have PDF uncertainties around 1.4\% and 2.0\%.

\item \textit{Strong coupling (\alpS)}: The uncertainties due to \alpS are estimated by using two additional PDF members corresponding to upper (0.120) and lower (0.116) values with respect to the central (0.118) \alpS value. The integrated cross sections for the \zgtoneb (\zgttwob) process have \alpS uncertainties around 0.2--0.3\% (0.1\%) in the dielectron, dimuon, and combined channels. For the \zgtoneb (\zgttwob) analysis, the uncertainty in the differential distributions for the dielectron, dimuon, and combined channels is 0.5--1.3\% (0.5--1.6\%), and the uncertainties in the normalized differential cross sections for these channels are in the range 0.5 to 1.1\% (0.5--1.3\%). The effect on \zgtoneb (\zgttwob) integrated cross section at the generator-level is 1.2\% (0.1\%). The \zgttwob and \zgtoneb \alpS uncertainties are summed in quadrature for the cross section ratios.
 
\end{itemize}

Table~\ref{tab:uncr_integralcross-sect} lists the uncertainties in the \zgtoneb and \zgttwob integrated cross sections. Tables~\ref{tab:uncert_range_norm_dilepton_fullRun2_xsec} and~\ref{tab:uncert_range_norm_dilepton_fullRun2} summarize the range of uncertainty in the differential and normalized differential cross section distributions for the \zgtoneb combined channel, respectively. For the \zgttwob measurements, Tables~\ref{tab:unc_range_zbb} and~\ref{tab:unc_range_norm_zbb} show the uncertainty in the differential and normalized differential cross sections, respectively. All systematic uncertainties are treated as 100\% correlated across the years. In summary, the \zgtoneb (\zgttwob) integrated cross sections have total experimental systematic uncertainties of 5.9 and 6.1\% (9.7 and 9.9\%), respectively, in the dimuon and combined channels. For the electron channel the uncertainty is a bit higher, up to 7.6\% (\zgtoneb) and 11.4\% (\zgttwob), owing to the electron selection uncertainty. For the \zgtoneb (\zgttwob) measured integrated cross section the statistical uncertainty is less than 1\% (5\%). 

\begin{table*}[htb!]
\centering
 \topcaption{Summary of the uncertainties (in percent) in the integrated cross sections for the dielectron, dimuon, and combined channels in the \zgtoneb and \zgttwob events.}
 \begin{scotch}{lccccccccc|}
                         &          &       \zgtoneb     &                &&&               &   \zgttwob             &\\
Uncertainty (\%)        & $\Pe\Pe$    & $\PGm\PGm$      & $\ell\ell$    && & $\Pe\Pe$    & $\PGm\PGm$  & $\ell\ell$                  \\
\hline
Statistical               & 1.0               & 0.7        & 0.6           &&&   7.7           &   5.9             &  4.6  \\
JES, JER                    & 2.7               & 3.0         & 2.9          && &   6.9           &   5.4             &  5.8\\
\PQb tagging/mistagging      & 3.0               & 2.9          & 2.9            &&&    5.4           &   6.0             &  5.8  \\
Unclustered energy of \ptmiss & 2.8               & 2.8         & 2.8             &&&       3.5       &        3.7        &   3.6  \\
Background estimation     & 2.2               & 2.0          & 2.1             &&&   2.3           &   2.4             &  2.4  \\
Pileup reweighting        & 2.4               & 1.7         & 1.9              &&&   2.9           &   2.1             &  2.4    \\
Electron selection        & 4.6               & \NA          & 1.5              &&&   4.3           &   \NA              &  1.4     \\
Luminosity                & 1.6               & 1.6           & 1.6            &&& 1.6               & 1.6               & 1.6 \\
Muon selection            & \NA                & 0.6         & 0.4              &&&   \NA            &   1.0             &  0.7    \\
Pileup jet identification & 0.3               & 0.3          & 0.3             &&&  0.6             &  0.7             & 0.7    \\ 
L1 prefiring              &0.3                 &0.2          &0.2              & &&   0.3           &   0.2             &  0.3  \\
$\mu_{\mathrm{R}}$ and $\mu_{\mathrm{F}}$ scales         & 2.6               & 2.9          & 2.1            &&&   2.5           &   2.3             &  2.5    \\
PDF                       & 0.4               & 0.3          & 0.3             &&&   0.3           &   0.3             &  0.3 \\
\alpS                  & 0.3               & 0.2          & 0.2         &&&    0.1            &           0.1             &  0.1 \\			
Total experimental       &7.6                &5.9          &6.1              &&&    11.4           &   9.7           & 9.9     \\
Total theoretical           &2.6                &2.9          &2.1            &&&  2.5             &2.3              & 2.5   \\ 
 \end{scotch}
\label{tab:uncr_integralcross-sect}
\end{table*}
 
\begin{table*}[htb!]
\centering
\topcaption{Summary of the uncertainties (in percent) in the differential cross section distributions for the combined channel in the \zgtoneb events.}
 \begin{scotch}{lcccccc}
Observable/Uncertainty (\%)      & \ZpT       &\bjetpT        &\bjetabseta    &\deltaPhi      &  \deltaY  & \deltaR  \\
\hline
Statistical                &1.0--5.2    & 1.0--12             &0.7--1.4       &0.7--2.2        &0.7--4.7    &1.0--5.3              \\
 JES, JER                   &1.0--9.0      & 2.0--7.2            &2.0--7.0    & 2.0--4.8          &2.0--6.4      &1.6--6.1        \\
\PQb tagging/mistagging        & 2.3--6.0   & 2.0--8.3            &2.6--4.3    &2.5--4.9        &2.7--5.5      &2.6--7.2   \\
Unclustered energy of \ptmiss &2.2--3.6      & 2.2--3.4      &0.5--1.8    & 2.5--3.8     & 2.7--3.0         & 2.3--3.8       \\
Background estimation       &1.6--3.6    &1.8--3.0         &1.9--3.2    &1.5--3.9         &2.0--3.4     &2.6--4.8      \\
Pileup reweighting             &1.0--2.2    &0.7--2.2         &1.8--2.4    & 1.8--2.6      &1.8--2.2       &1.5--2.5    \\
Model dependency             &0.5--1.4   &0.5--1.2         &0.5         &0.5--1.5      &0.5--0.6        & 0.5--1.1    \\
Electron selection           & 1.3--2.8   &1.2--1.9        &1.4--1.8    &1.3--1.9      &1.5--2.4        &1.2--2.2      \\
Muon selection               & 0.5--1.0      &0.5--1.0         &0.5         & 0.5          &0.5             &0.5--0.6      \\
Pileup jet identification    & 0.5--0.6    &0.5            &0.5--0.9    &0.5           &0.5--0.6       &0.5    \\
L1 prefiring                 & 0.5         &0.5            & 0.5--0.7    &0.5          & 0.5--0.9      &0.5     \\
$\mu_{\mathrm{R}}$ and $\mu_{\mathrm{F}}$ scales & 1.0--7.2     &1.0--10.0     &0.7--3.3   &1.5--2.7        &1.0--2.8       &1.5--4.0  \\
PDF                        & 0.5--2.9      &0.5--2.1       &1.5--2.8    & 0.5--2.1       &0.5--1.6      &0.5--4.2  \\
\alpS                       & 0.5--1.0        &0.5--0.7       &0.5         &0.5--1.1        &0.5           &0.5--1.3  \\ 
\end{scotch}
\label{tab:uncert_range_norm_dilepton_fullRun2_xsec}
\end{table*}

\begin{table*}[htb!]
 \centering
 \topcaption{Summary of the uncertainties (in percent) in the normalized differential distributions for the combined channel in the \zgtoneb events.}
 \cmsTable{
 \begin{scotch}{lcccccc}
    Observable/Uncertainty (\%)     & \ZpT      &\bjetpT     &\bjetabseta     &\deltaPhi   &  \deltaY & \deltaR \\
   \hline
   Statistical        &1.0--5.2             & 1.0--12         &0.7--1.4       &1.0--2.1    &0.7--4.7   &0.8--3.3   \\
   JES, JER           &1.0--5.8               & 1.0--6.7       &0.5--4.5    & 0.5--2.1     &0.5--4.2  &0.5--4.2      \\
   \PQb tagging/mistagging & 0.5--3.9          &0.5--5.6       &0.5--1.7    &0.5--2.9         &0.5--3.7      &0.5--1.7       \\
   Unclustered energy of \ptmiss & 0.5--0.7         & 0.5--0.7      &0.5--1.1   &0.5--1.0          &0.5       &0.5--1.0      \\
   Background estimation & 0.5--1.9         &0.6--1.7        &0.5--1.4    &0.5--2.0       &0.5--2.4  &0.5--3.3      \\
   Pileup reweighting   &0.5--0.9            &0.5--1.3         &0.5   & 0.5--0.7           &0.5        &0.5--0.8      \\
   Model dependency    &0.5--1.4            &0.5--0.8          &0.5     &0.5--1.2          &0.5        & 0.5--0.9          \\
   Electron selection  & 0.5--1.3           & 0.5           &0.5       &0.5               &0.5--0.9   &0.5--0.6       \\
   Muon selection      & 0.5--1.0             &0.5            &0.5      & 0.5              & 0.5      &0.5        \\
   Pileup jet identification & 0.5         &0.5            &0.5      &0.5                   &0.5       &0.5     \\
   L1 prefiring             & 0.5          &0.5            &0.5      &0.5                  &0.5       &0.5           \\
   $\mu_{\mathrm{R}}$ and $\mu_{\mathrm{F}}$ scales  & 1.0--7.0  &0.5--9.4   &0.5--2.0     &1.0--2.1          &0.5--2.1    &0.5--3.7       \\
   PDF                       &0.5--2.0      & 0.5--1.6      &0.5--0.7   & 0.5--1.8            &0.5--1.6    &0.5--3.3   \\ 
   \alpS                   &0.5--1.0         & 0.5--0.7     & 0.5       &0.5--0.9             &  0.5        &0.5--1.1       \\
  
\end{scotch}
}
\label{tab:uncert_range_norm_dilepton_fullRun2}
\end{table*}

\begin{table*}[htb!]
\topcaption{Summary of the uncertainties (in percent) in the differential cross section distributions for the combined channel in the \zgttwob events. The symbols, $\PQb1$ and $\PQb2$, stand for leading and subleading \PQb jets, respectively.}
\cmsTable{
\begin{scotch}{lccccccccc}
Observable/Uncertainty (\%) & $\pt^{\PQb1}$ & $\pt^{\PQb2}$ & $\abs{\eta}^{\PQb1}$ & $\pt^{\PZ}$ & $\Delta R_{\PQb\PQb}$ & $\Delta R^\text{min}_{\PZ\PQb\PQb}$ & $A_{\PZ\PQb\PQb}$ & $m_{\PZ\PQb\PQb}$ & $m_{\PQb\PQb}$\\
\hline
Statistical & 8.6--15.5 & 7.0--37.6 & 6.2--11.9 & 6.2--25.3 & 5.6--29.2 & 5.3--43.5 & 4.5--25.0 & 8.1--22.1 & 11.3--25.9 \\
JES, JER & 6.0--12.8 & 5.1--14.3 & 6.8--8.2 & 5.6--14.8 & 5.9--14.9 & 4.5--23.8 & 4.3--13.2 & 4.8--13.7 & 4.2--17.5 \\
\PQb tagging/mistagging & 4.4--7.7 & 4.3--9.2 & 4.3--6.2 & 4.1--13.7 & 3.5--6.4 & 3.8--14.5 & 5.3--7.3 & 4.2--7.2 & 4.5--6.6 \\
Unclustered energy of \ptmiss & 1.3--6.8 & 2.1--8.0 & 2.7--4.2 & 3.1--6.8 & 1.1--5.0 & 0.7--4.8 & 2.1--4.8 & 2.1--5.1 & 1.3--8.4 \\
Background simulation & 0.6--3.7 & 0.8--4.9 & 1.0--2.9 & 0.6--3.8 & 0.5--3.0 & 0.5--3.8 & 1.9--3.1 & 0.6--4.1 & 1.1--3.5 \\
Pileup reweighting & 1.4--4.1 & 1.6--6.8 & 1.7--2.7 & 1.6--3.3 & 1.9--3.1 & 0.9--3.9 & 1.7--4.3 & 1.7--3.2 & 1.5--3.5 \\
Electron selection & 0.9--1.7 & 1.1--2.0 & 1.2--1.5 & 0.9--1.7 & 0.7--1.5 & 0.5--1.8 & 1.1--1.6 & 1.0--1.8 & 0.9--1.6 \\
Muon selection & 0.5--0.8 & 0.5--0.9 & 0.7--3.3 & 0.4--0.9 & 0.5--0.8 & 0.5--0.8 & 0.6--0.8 & 0.5--0.9 & 0.6--0.8 \\
Pileup jet identification & 0.6--1.2 & 0.5--1.0 & 0.8--0.9 & 0.7--0.9 & 0.8--1.1 & 0.8--0.9 & 0.7--0.8 & 0.7--1.1 & 0.6--1.0 \\
$\mu_{\mathrm{R}}$ and $\mu_{\mathrm{F}}$ scales & 0.5--3.2 & 0.7--4.6 & 1.8--4.7 & 0.7--4.5 & 0.6--4.1 & 0.7--4.7 & 1.3--4.9 & 1.7--3.8 & 1.1--4.8 \\
PDF & 0.5--0.8 & 0.5--1.4 & 0.5--0.7 & 0.5--1.1 & 0.5--2.5 & 0.5--4.0 & 0.5--0.9 & 0.5--0.8 & 0.5--0.9 \\
\alpS & 0.5 & 0.5 & 0.5 & 0.5 & 0.5--1.6 & 0.5--0.9 & 0.5 & 0.5 & 0.5 \\
\end{scotch}
}
\label{tab:unc_range_zbb}
\end{table*}

\begin{table*}[htb!]
\topcaption{Summary of the uncertainties (in percent) in the normalized differential distributions for the combined channel in the \zgttwob events. The symbols, $\PQb1$ and $\PQb2$, stand for leading and subleading \PQb jets, respectively.}
\cmsTable{
\begin{scotch}{lccccccccc}
Observable/Uncertainty (\%) & $\pt^{\PQb1}$ & $\pt^{\PQb2}$ & $\abs{\eta}^{\PQb1}$ & $\pt^{\PZ}$ & $\Delta R_{\PQb\PQb}$ & $\Delta R^\text{min}_{\PZ\PQb\PQb}$ & $A_{\PZ\PQb\PQb}$ & $m_{\PZ\PQb\PQb}$ & $m_{\PQb\PQb}$\\
\hline
Statistical & 9.7--16.2 & 8.3--37.9 & 7.7--12.8 & 7.7--25.7 & 7.2--29.6 & 7.0--43.7 & 6.4--25.4 & 9.3--22.6 & 12.2--26.3 \\
JES, JER & 3.2--7.7 & 1.6--9.4 & 0.8--1.3 & 0.9--7.9 & 0.8--7.9 & 1.0--17.5 & 0.9--6.1 & 2.2--8.1 & 3.5--12.5 \\
\PQb tagging/mistagging & 0.5--2.2 & 0.7--3.8 & 0.6--1.7 & 0.5--8.4 & 0.5--2.4 & 0.5--9.5 & 0.5--1.6 & 0.5--1.7 & 0.5--1.3 \\
Unclustered energy of \ptmiss & 0.5--3.1 & 0.8--7.1 & 0.5--0.9 & 0.5--3.0 & 0.5--2.5 & 0.5--6.1 & 0.5--3.9 & 0.7--3.1 & 0.5--4.6 \\
Background simulation & 0.5--1.8 & 0.5--2.7 & 0.5--1.4 & 0.5--1.9 & 0.5--2.1 & 0.5--2.3 & 0.5--0.7 & 0.5--1.8 & 0.5--1.3 \\
Pileup reweighting & 0.5--1.8 & 0.5--4.3 & 0.5--0.8 & 0.5--2.7 & 0.5--0.8 & 0.5--1.7 & 0.5--1.7 & 0.5--1.1 & 0.5--1.3 \\
Electron selection & 0.5 & 0.5--0.7 & 0.5 & 0.5 & 0.5--0.6 & 0.5--1.0 & 0.5 & 0.5 & 0.5 \\
Muon selection & 0.5 & 0.5 & 0.9--2.3 & 0.5 & 0.5 & 0.5 & 0.5 & 0.5 & 0.5 \\
Pileup jet identification & 0.5 & 0.5 & 0.5 & 0.5 & 0.5 & 0.5 & 0.5 & 0.5 & 0.5 \\
$\mu_{\mathrm{R}}$ and $\mu_{\mathrm{F}}$ scales & 0.5--2.9 & 0.6--4.6 & 1.4--3.9 & 1.4--3.8 & 1.3--5.0 & 1.1--4.8 & 1.7--4.6 & 2.0--4.7 & 1.3--4.9 \\
PDF & 0.5--0.6 & 0.7--1.3 & 0.5--0.6 & 0.5--1.0 & 0.5--2.1 & 2.6--3.2 & 0.5--0.9 & 0.5--0.8 & 0.5--0.8 \\
\alpS & 0.5 & 0.5 & 0.5 & 0.5 & 0.5--1.3 & 0.5--0.6 & 0.5 & 0.5 & 0.5 \\
\end{scotch}
}
\label{tab:unc_range_norm_zbb}
\end{table*}

\clearpage
\section{Results}
The measured \zgtoneb and \zgttwob integrated cross sections in the fiducial region described in Table~\ref{table:1}, are $6.52\pm 0.04\stat\pm 0.40\syst\pm 0.14\thy$\unit{pb} and $0.65\pm 0.03\stat\pm 0.07\syst\pm 0.02\thy$\unit{pb} in the combined channel, respectively. The measurements are consistent for the dielectron, dimuon, and combined channels as reported in Table~\ref{tab:integrated_cross_Sec_full}, therefore the differential and normalized differential distributions are shown only for the combined channel. 
The measured cross sections are compared with predictions from \MGfive obtained with two settings corresponding to 2016 and 2017--2018 data-taking periods. The 2017--2018 settings use more up-to-date \MGfive versions (2.6.0 and 2.4.2 for NLO and LO, respectively), PDFs (NNPDF 3.1), and underlying event tune (CP5), and are labeled with NNPDF 3.1, CP5 in the following. The 2016 data-taking period uses earlier versions of the \MGfive (2.3.2.2 and 2.2.2 for NLO and LO, respectively), PDFs (NNPDF 3.0), and underlying event tune (CUETP8M1), and are labeled with NNPDF 3.0, CUETP8M1. In the \zgtoneb final state, the measured integrated cross section values are well described by \MGfive LO for both settings, whereas \MGfive (NLO, NNPDF 3.1, CP5) and \MGfive (NLO, NNPDF 3.0, CUETP8M1) overestimate data by $\approx$10\ and $\approx$18\%, respectively. The \SHERPA simulation overestimates the measured cross section values by $\approx$24\%. In the \zgttwob final state, the measured integrated cross section is in good agreement with \MGfive (LO). The \MGfive (NLO, NNPDF 3.1, CP5), \MGfive (NLO, NNPDF 3.0, CUETP8M1), and \SHERPA predictions overestimate the measured cross section values by 29, 38, and 21\%, respectively. 
The measured value of the cross section ratio of the \zgttwob and \zgtoneb for the combined channel is $0.100\pm 0.005\stat\pm 0.007\syst\pm 0.003\thy$ and, as expected, has smaller uncertainties. This measurement is in agreement with the \MGfive (LO, NNPDF 3.0, CUETP8M1) and \SHERPA values given the large uncertainties associated with the predictions. Table~\ref{tab:integrated_cross_Sec_full} summarizes the measured and predicted cross sections for the \zgtoneb and \zgttwob processes. 
 \begin{table*}[htb!]
 \centering
 \topcaption{Measured and predicted cross sections (in\unit{pb}) for the \zgtoneb and \zgttwob final states. The cross section ratios between the \zgttwob and \zgtoneb are shown in the last three rows for the dielectron, dimuon, and combined channels. In the measured results the first, second, and third uncertainties correspond to the statistical, systematic, and theoretical sources, respectively. The \MGfive (NLO) predictions include theoretical uncertainties (PDF, and renormalization and factorization scales).}

\label{tab:integrated_cross_Sec_full}
 \cmsTable{
     \begin{scotch}{l c  c  c c c  c c }
       & Channel &  Measured    & \MGfive              & \MGfive            & \MGfive        & \MGfive     & \SHERPA  \\    
       &           &            & LO                    & LO                 & NLO             & NLO         &       \\    
       &         &             &  NNPDF 3.0         &  NNPDF 3.1             & NNPDF 3.0      & NNPDF 3.1 & \\
       &         &             &   CUETP8M1               &  CP5             & CUETP8M1            &  CP5 & \\
\hline
\zgtoneb &  $\Pe\Pe$  & 6.45 $\pm$ 0.06 $\pm$ 0.49 $\pm$ 0.17  & 6.25 & 6.33     & 7.86 $\pm$ 0.52 & 7.05 $\pm$ 0.48      & 8.05 \\
          &  $\PGm\PGm$  & 6.55 $\pm$ 0.05 $\pm$ 0.39 $\pm$ 0.19 & 6.26 & 6.34    &7.86 $\pm$ 0.51    & 7.02 $\pm$ 0.47    &  7.98 \\
 & $\ell\ell$ & 6.52  $\pm$ 0.04 $\pm$ 0.40 $\pm$ 0.14           &6.25 &6.34        & 7.86 $\pm$ 0.51 & 7.03 $\pm$ 0.47      & 8.02 \\
\zgttwob & $\Pe\Pe$ & 0.66 $\pm$ 0.05 $\pm$ 0.07 $\pm$ 0.02     &0.62 & 0.72       &0.89 $\pm$ 0.08    & 0.77 $\pm$ 0.07      & 0.84\\
& $\PGm\PGm$ & 0.65 $\pm$ 0.04 $\pm$ 0.06 $\pm$ 0.02            &0.64 & 0.71       & 0.91 $\pm$ 0.09  & 0.77 $\pm$ 0.07  & 0.84\\
& $\ell\ell$ & 0.65 $\pm$ 0.03 $\pm$ 0.07 $\pm$ 0.02            &0.63 & 0.71       &0.90 $\pm$ 0.09  &0.77 $\pm$ 0.07      & 0.84 \\
Ratio & $\Pe\Pe$   &0.102 $\pm$ 0.008 $\pm$ 0.008 $\pm$ 0.004  &0.100 & 0.113        & 0.113 $\pm$ 0.016 &0.110 $\pm$ 0.013   &0.104 \\
 & $\PGm\PGm$   &  0.100 $\pm$ 0.006 $\pm$ 0.006 $\pm$ 0.004  &0.103 & 0.112       & 0.116 $\pm$ 0.016  &0.110 $\pm$ 0.013   &0.105 \\
 & $\ell\ell$   &  0.100 $\pm$ 0.005 $\pm$ 0.007 $\pm$ 0.003  &0.102 & 0.112     & 0.114 $\pm$ 0.016     &0.110 $\pm$ 0.013   &0.105 \\

 \end{scotch}
    }
 \end{table*}
 
\begin{figure*}[ht!]
\centering
{\includegraphics[width=0.44\textwidth]{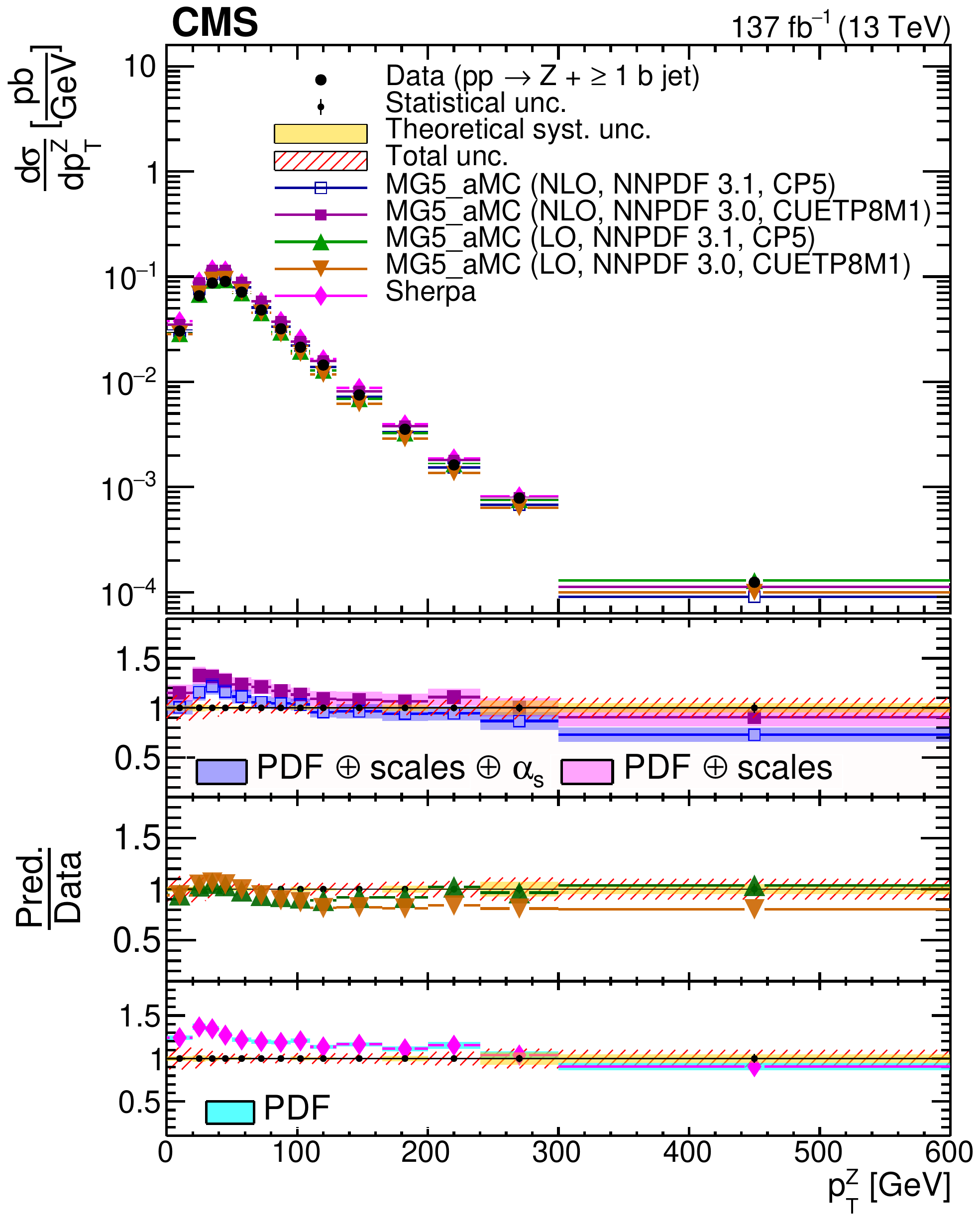}\label{fig:cross_dileptonpt_xsec}}
{\includegraphics[width=0.44\textwidth]{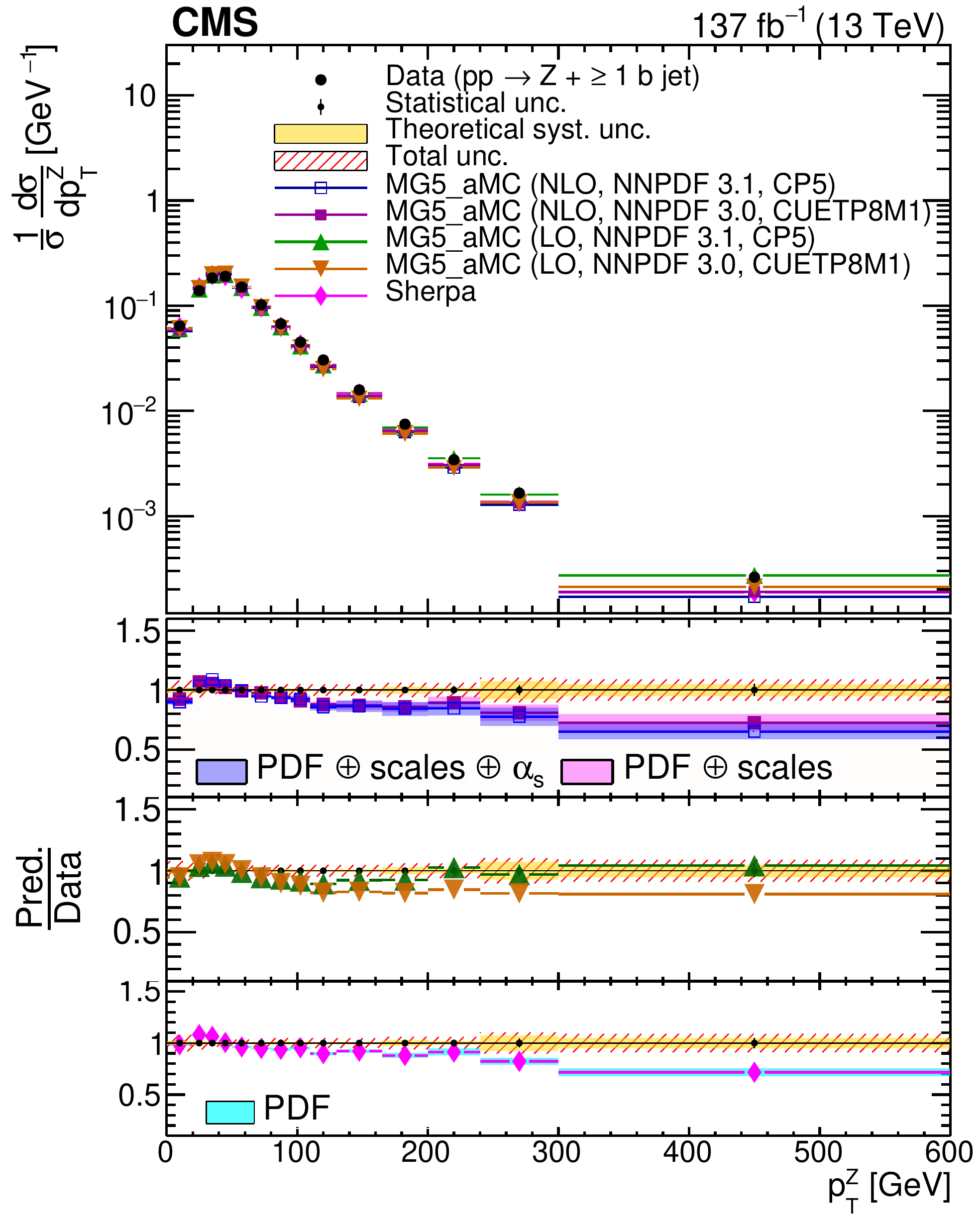}\label{fig:cross_dileptonpt_norm}}
\caption{ (\cmsLeft) Differential cross section and the (\cmsRight) normalized differential cross section distributions as a function of \PZ transverse momenta for \zgtoneb events. The uncertainties in the predictions are shown as colored bands in the bottom panel around the theoretical predictions including statistical, PDF, scale, and \alpS uncertainties for the \MGfive (NLO, NNPDF 3.1, CP5), \MGfive(NLO, NNPDF 3.0, CUETP8M1) and the statistical and PDF uncertainties for the \SHERPA predictions. The statistical, theoretical, and total uncertainties in data are indicated by the vertical bars, yellow, and hatched bands centered at 1, respectively. The same description applies to all of the remaining figures.} \label{fig:cross_dilepton_pt}
\end{figure*}

\begin{figure*}[htbp]
\centering
{\includegraphics[width=0.44\textwidth]{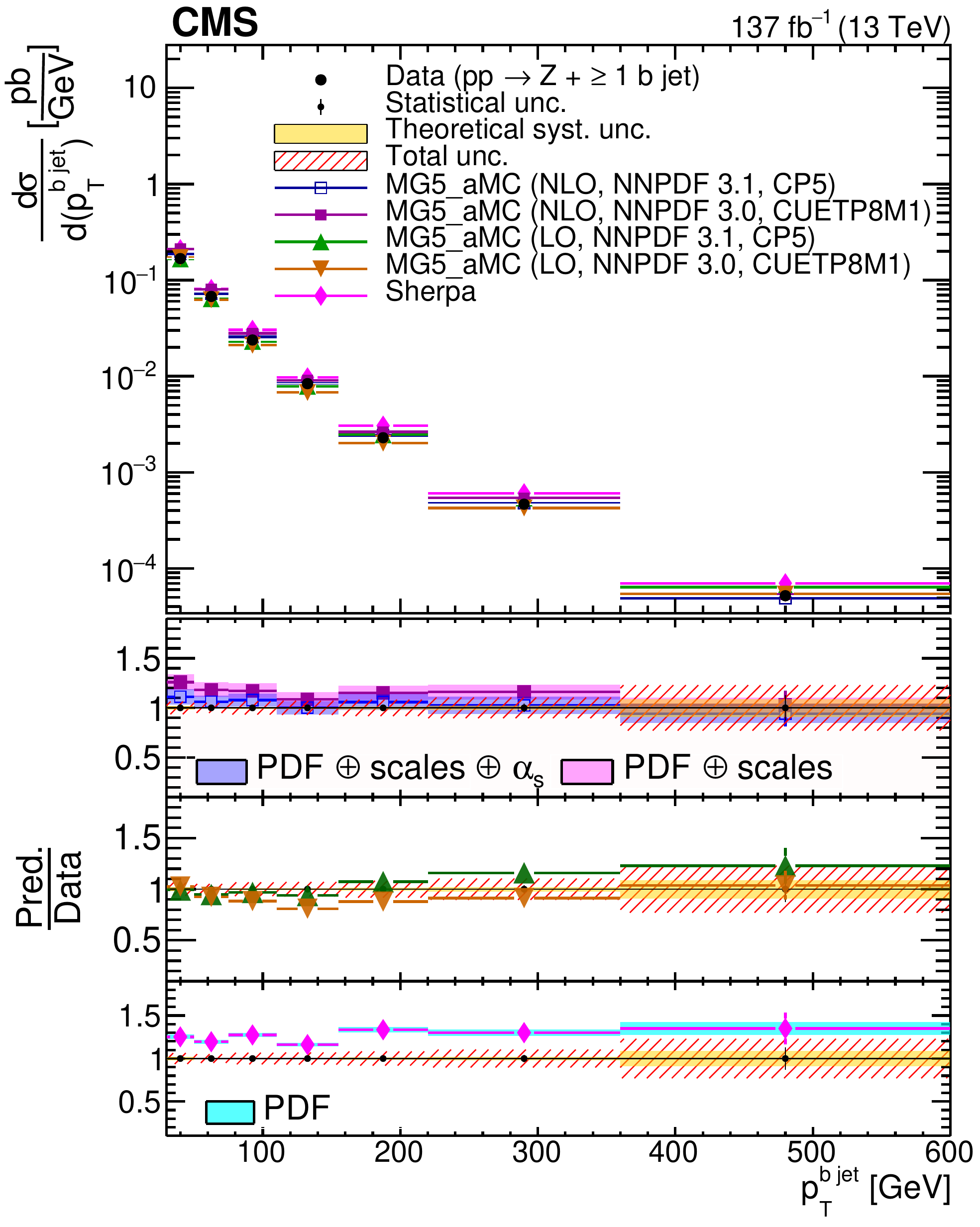}\label{fig:cross_dileptonbjetpt_xsec}}
{\includegraphics[width=0.44\textwidth]{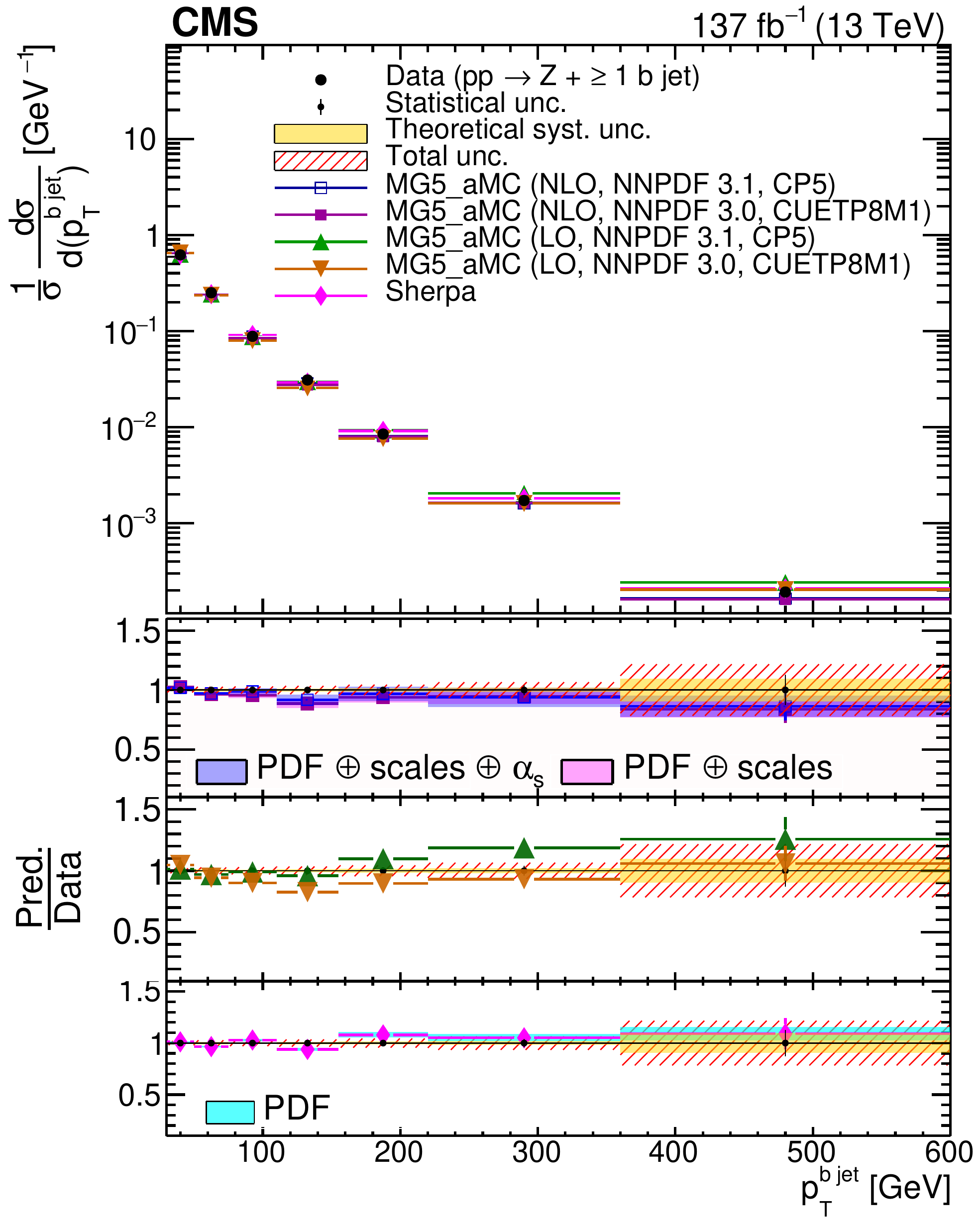}\label{fig:cross_dileptonbjetpt_norm}}
\caption{(\cmsLeft) Differential cross section and the (\cmsRight) normalized differential cross section distributions as a function of \PQb jet transverse momenta for \zgtoneb events.}
\label{fig:cross_dileptonbjetpt_pt}
\end{figure*}

\begin{figure*}[ht!]
\centering
{\includegraphics[width=0.44\textwidth]{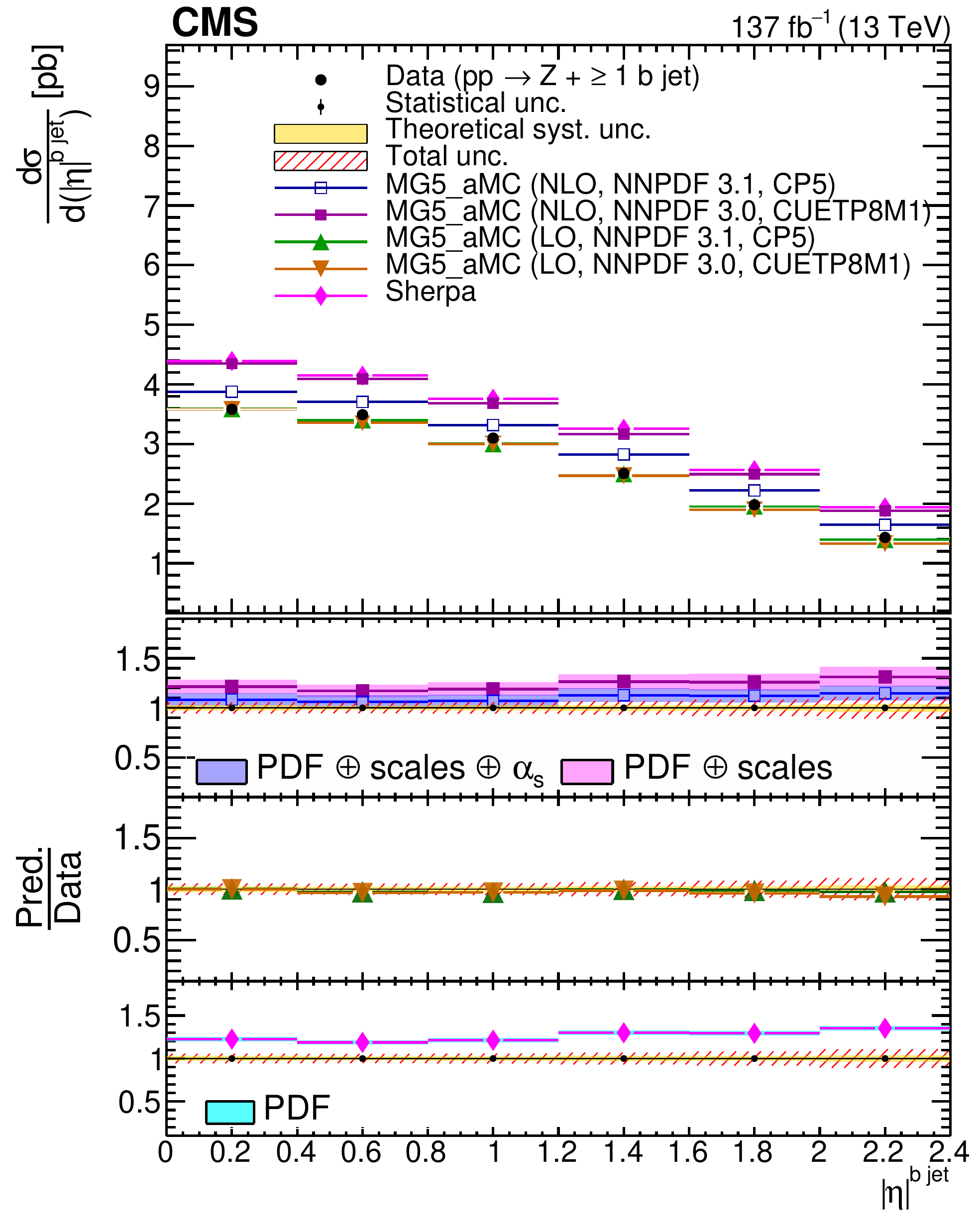}\label{fig:cross_dileptonbjetabseta_xsec}}
{\includegraphics[width=0.44\textwidth]{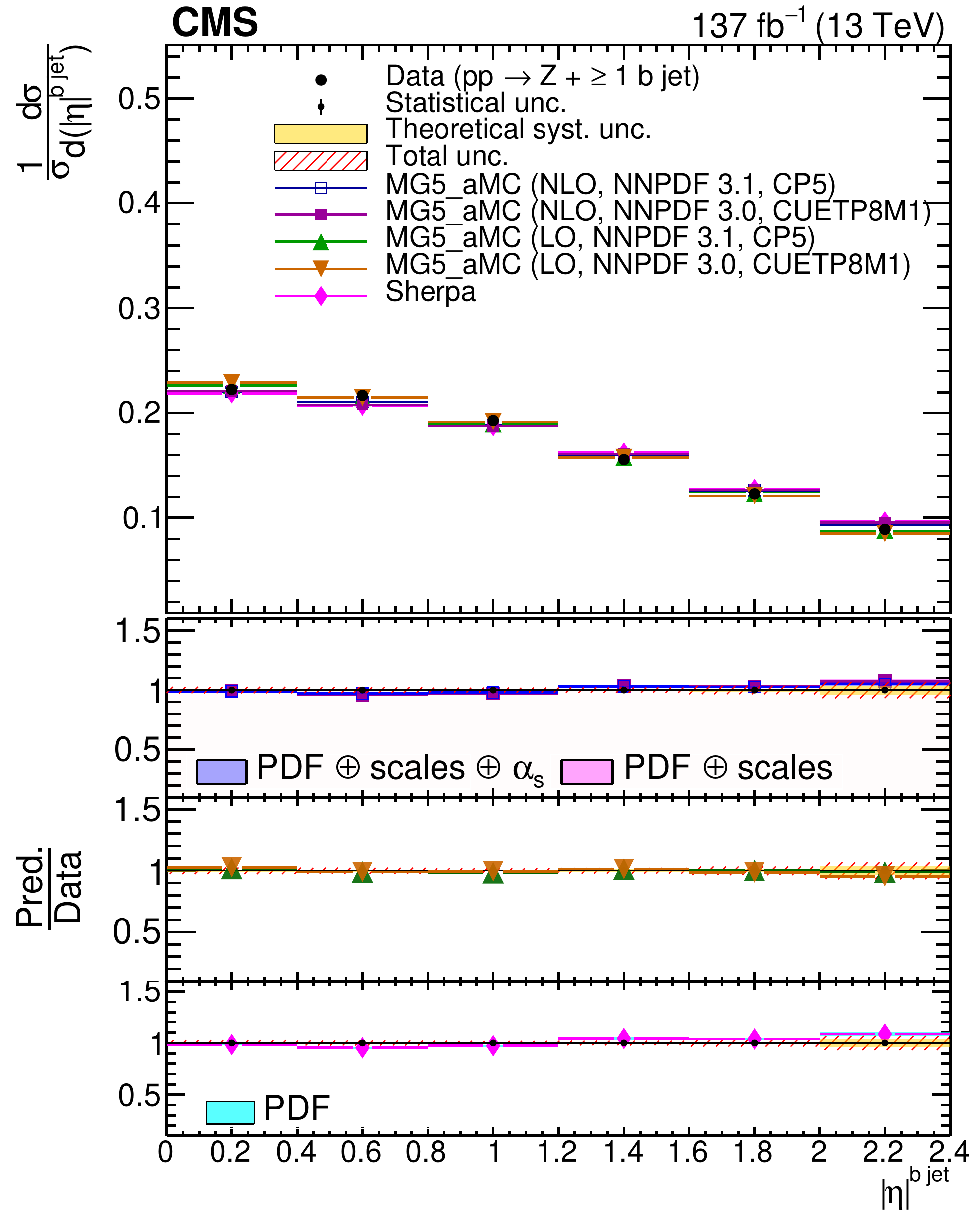}\label{fig:cross_dileptonbjetabseta_norm}}
\caption{(\cmsLeft) Differential cross section and the (\cmsRight) normalized differential cross section distributions as functions of \PQb jet absolute pseudorapidity for \zgtoneb events.}
\label{fig:cross_dileptonbjet_eta}
\end{figure*}

\begin{figure*}[ht!]
\centering
{\includegraphics[width=0.44\textwidth]{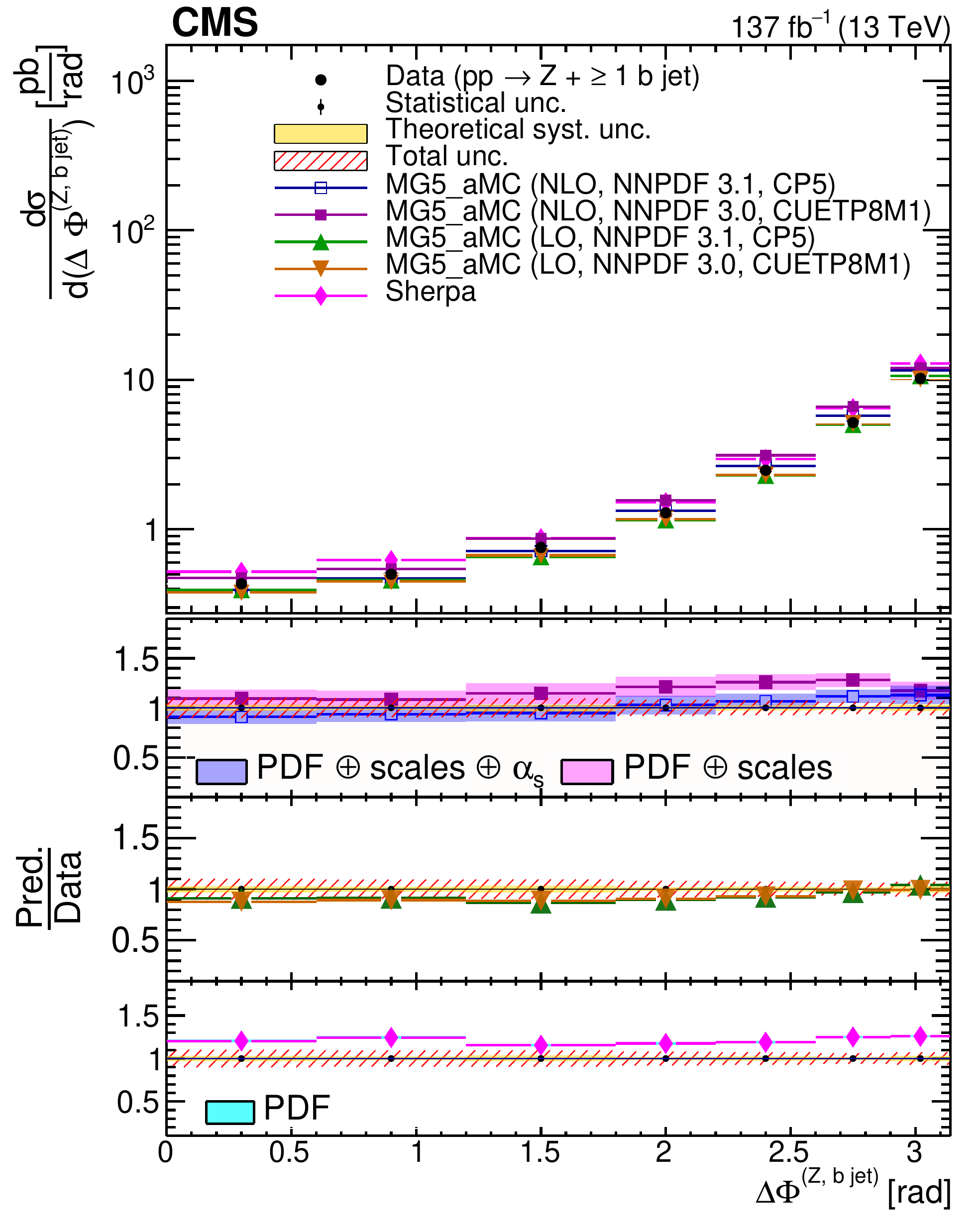}\label{fig:cross_dileptondeltaPhi_xsec}}
{\includegraphics[width=0.44\textwidth]{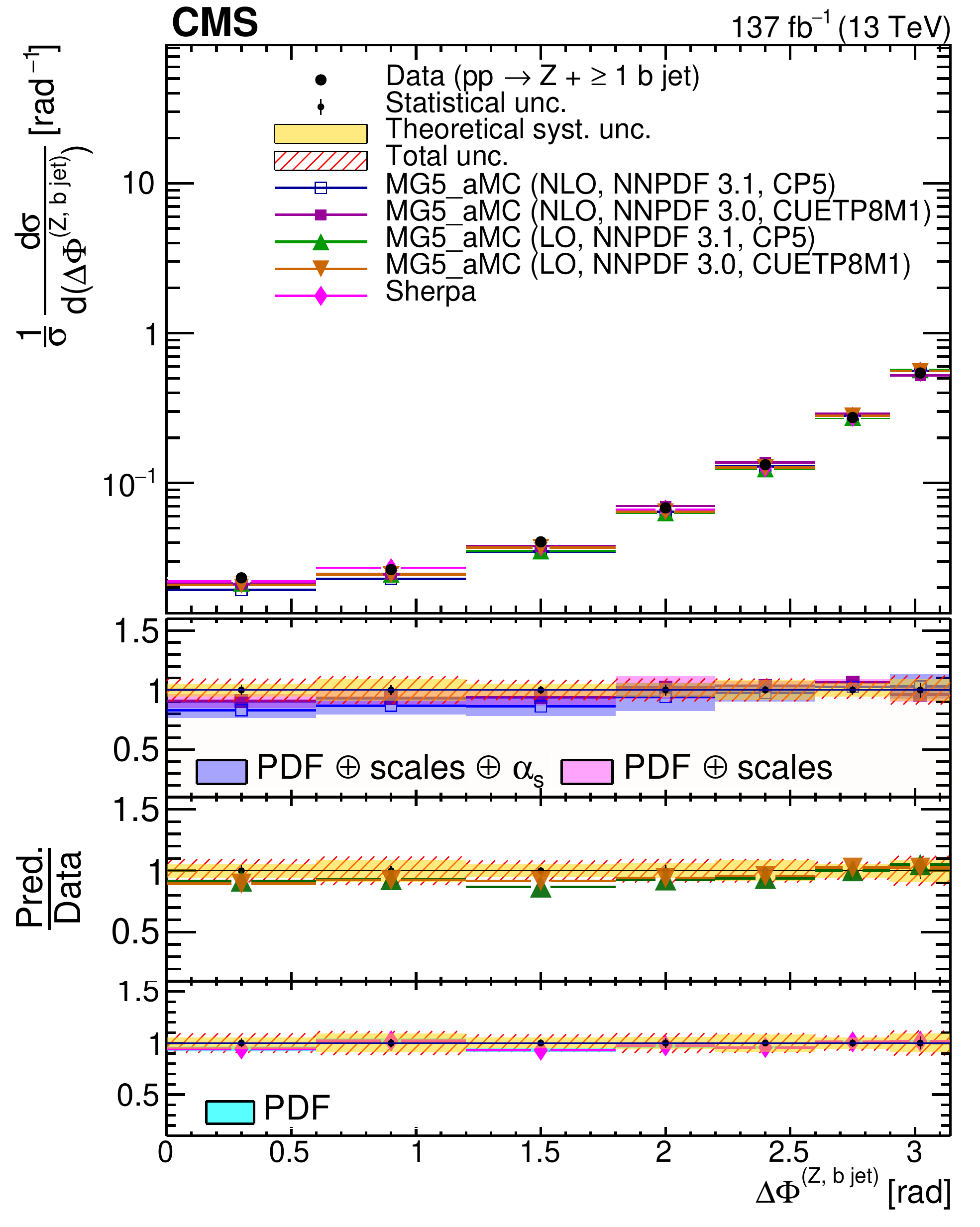}\label{fig:cross_dileptondeltaPhi_norm}}
\caption{(\cmsLeft) Differential cross section and the (\cmsRight) normalized differential cross section distributions as functions of \deltaPhi between the \PZ boson and the leading \PQb jet for \zgtoneb events.}
 \label{fig:cross_dilepton_delphi}
\end{figure*}

\begin{figure*}[htb]
\centering
{\includegraphics[width=0.44\textwidth]{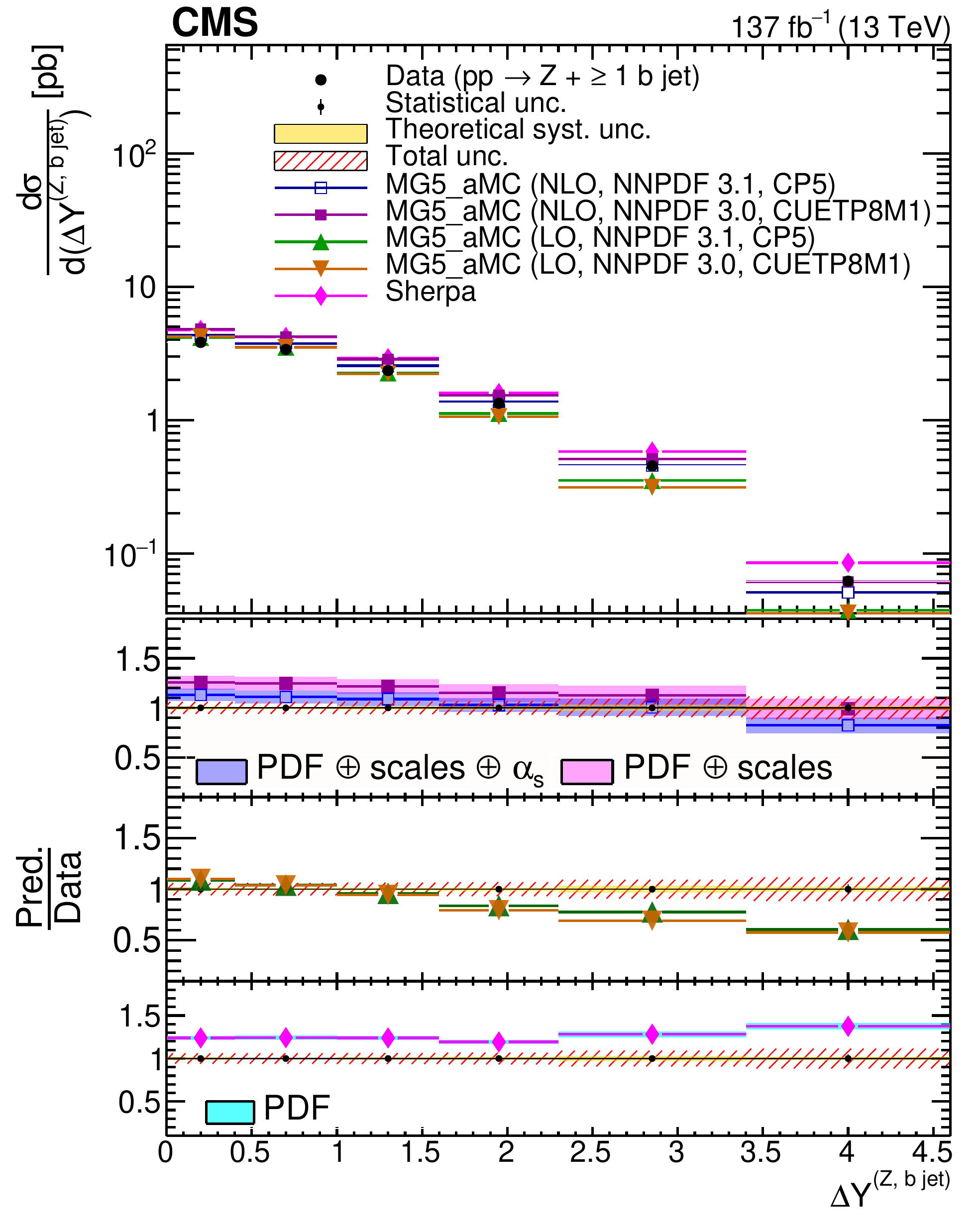}\label{fig:cross_dileptondeltaY_xsec}}
{\includegraphics[width=0.44\textwidth]{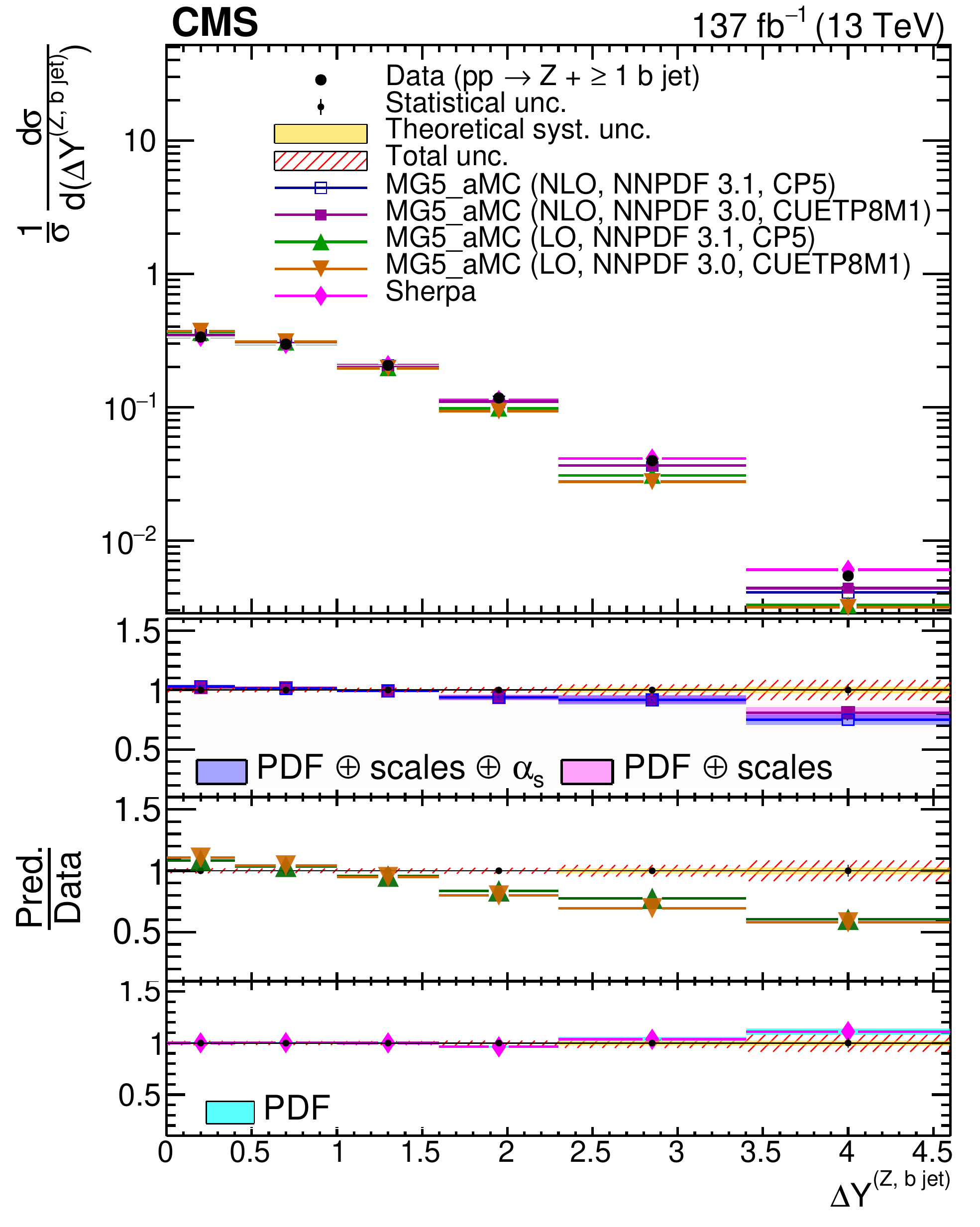}\label{fig:cross_dileptondeltaY_norm}}
\caption{(\cmsLeft) Differential cross section and the (\cmsRight) normalized differential cross section distributions as functions of \deltaY between the \PZ boson and the leading \PQb jet for the \zgtoneb events.} 
\label{fig:cross_dilepton_phiY}
\end{figure*}

\begin{figure*}[htb]
\centering
{\includegraphics[width=0.44\textwidth]{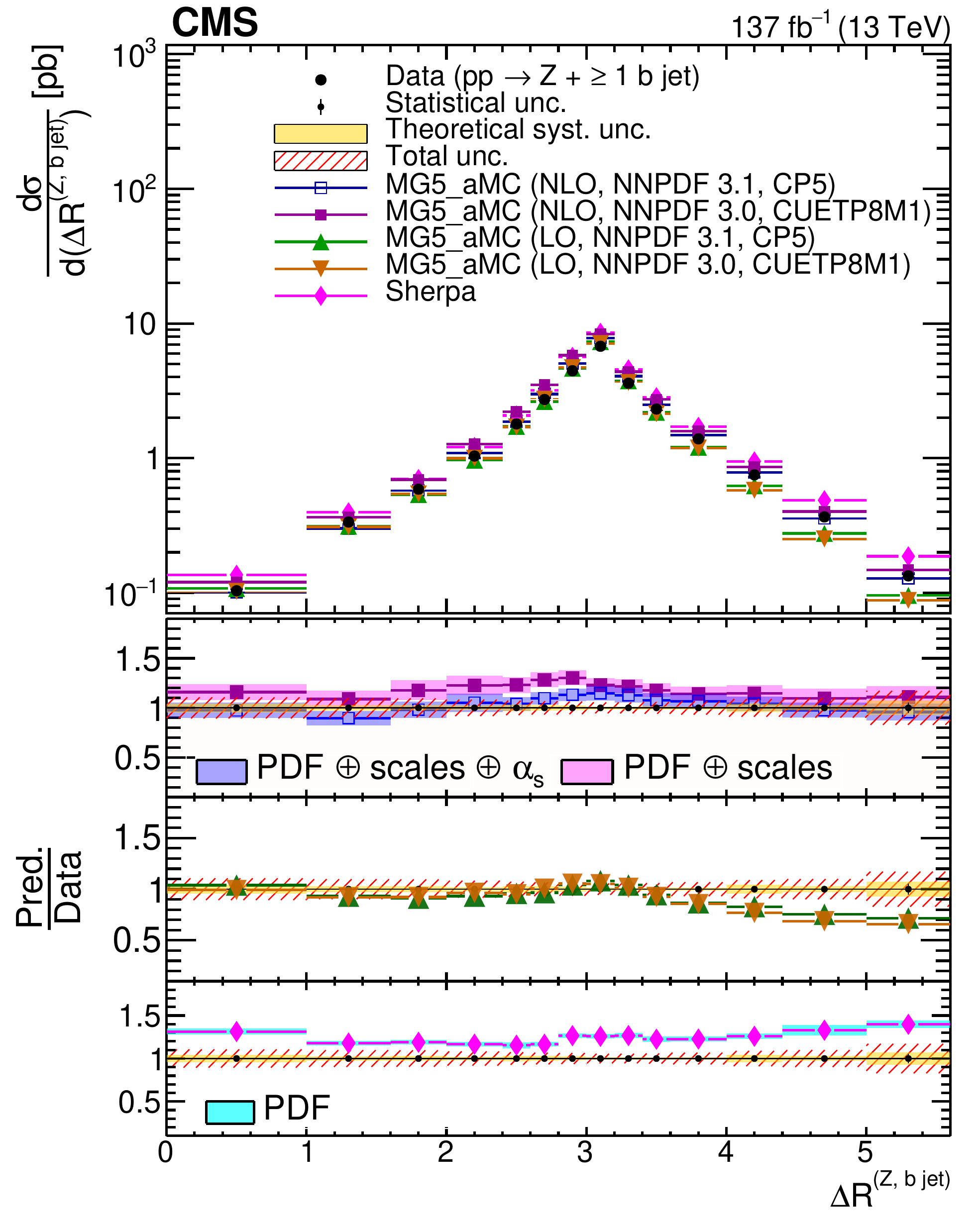}\label{fig:cross_dileptondetaR_xsec}}
{\includegraphics[width=0.44\textwidth]{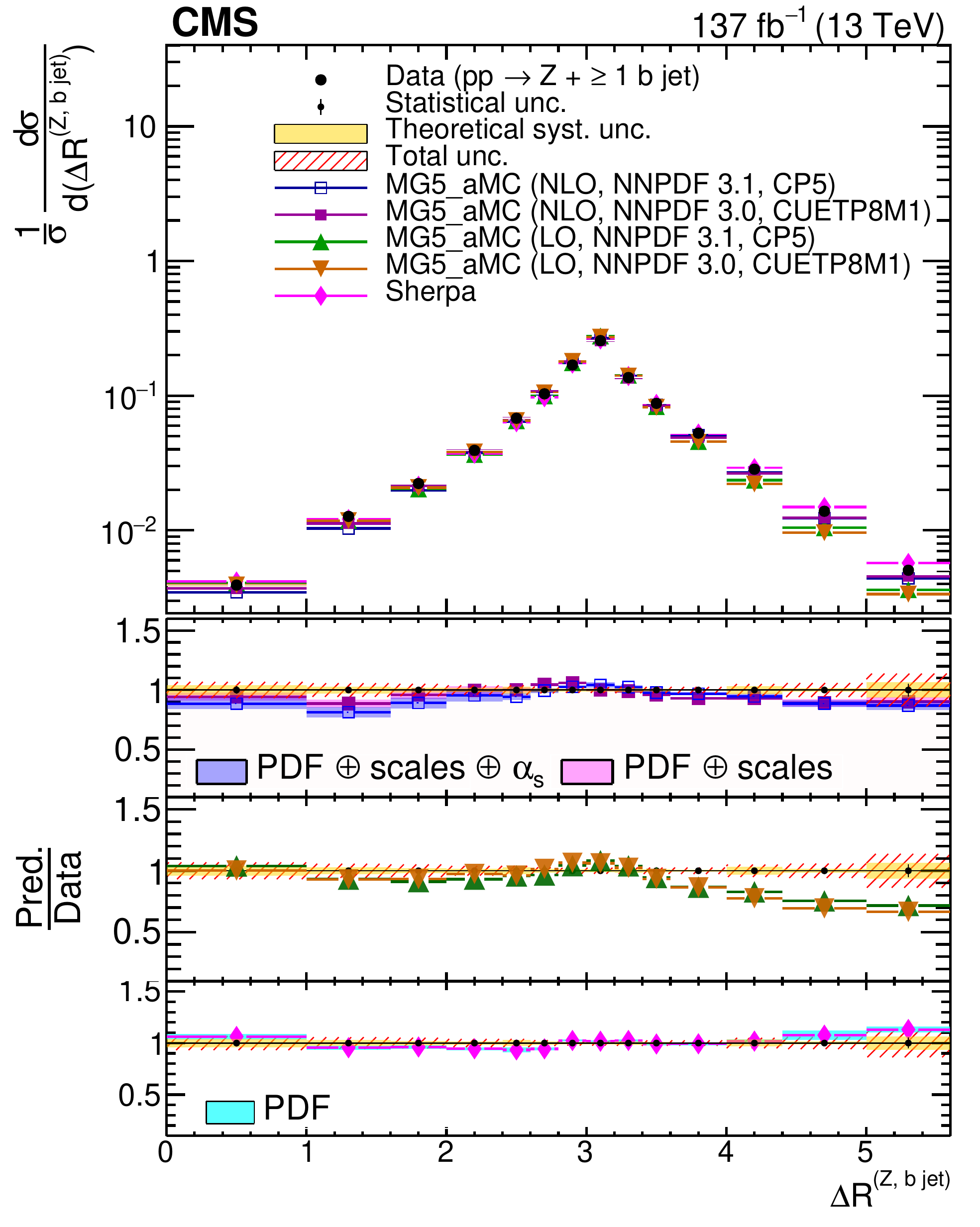}\label{fig:cross_dileptondetaR_norm}}
\caption{(\cmsLeft) Differential cross section and the (\cmsRight) normalized differential cross section distributions as functions of \deltaR between the \PZ boson and the leading \PQb jet for the \zgtoneb events.}
\label{fig:cross_dilepton_delR}
\end{figure*}

\begin{figure*}[htb]
\centering
{\includegraphics[width=0.44\textwidth]{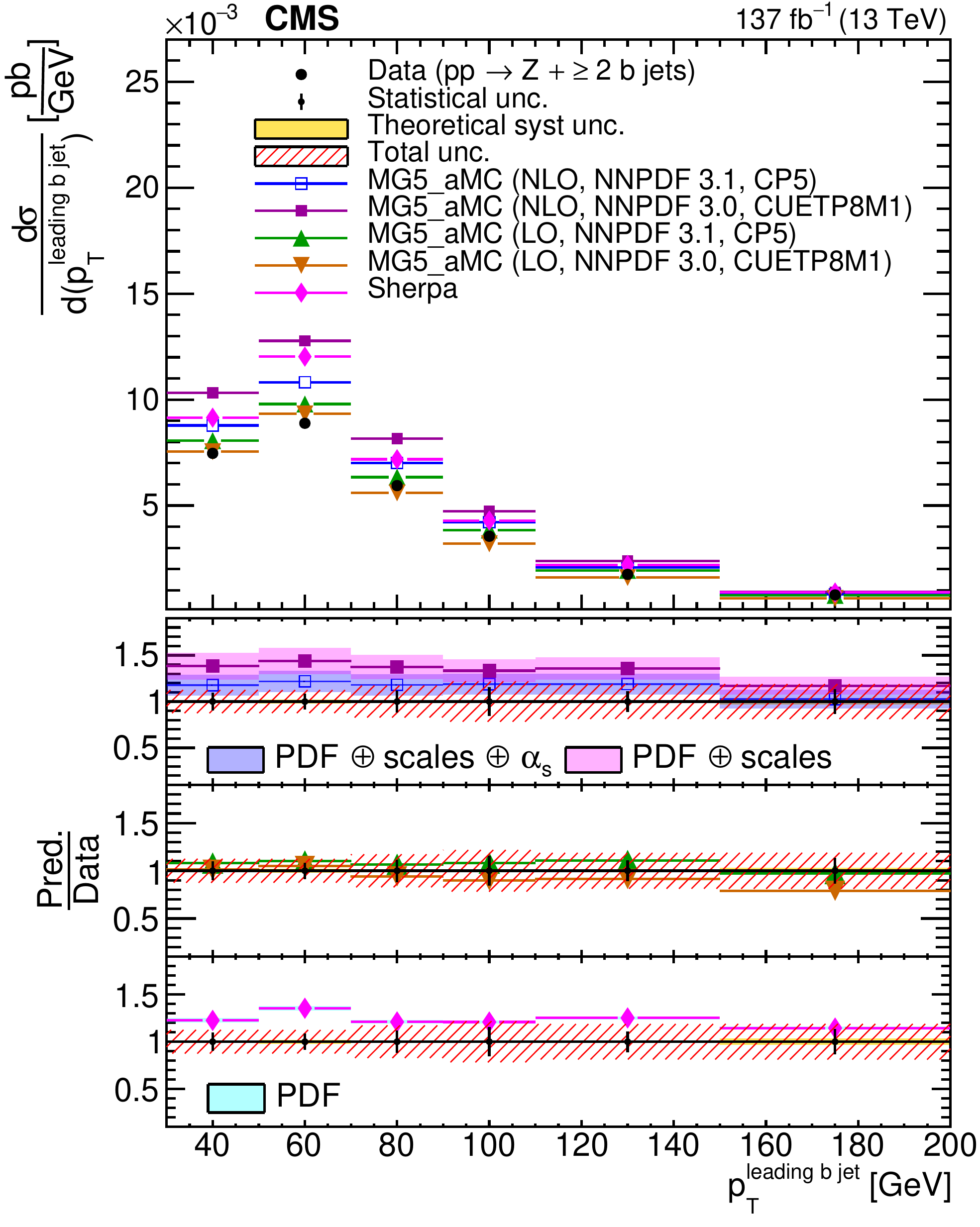}}
{\includegraphics[width=0.44\textwidth]{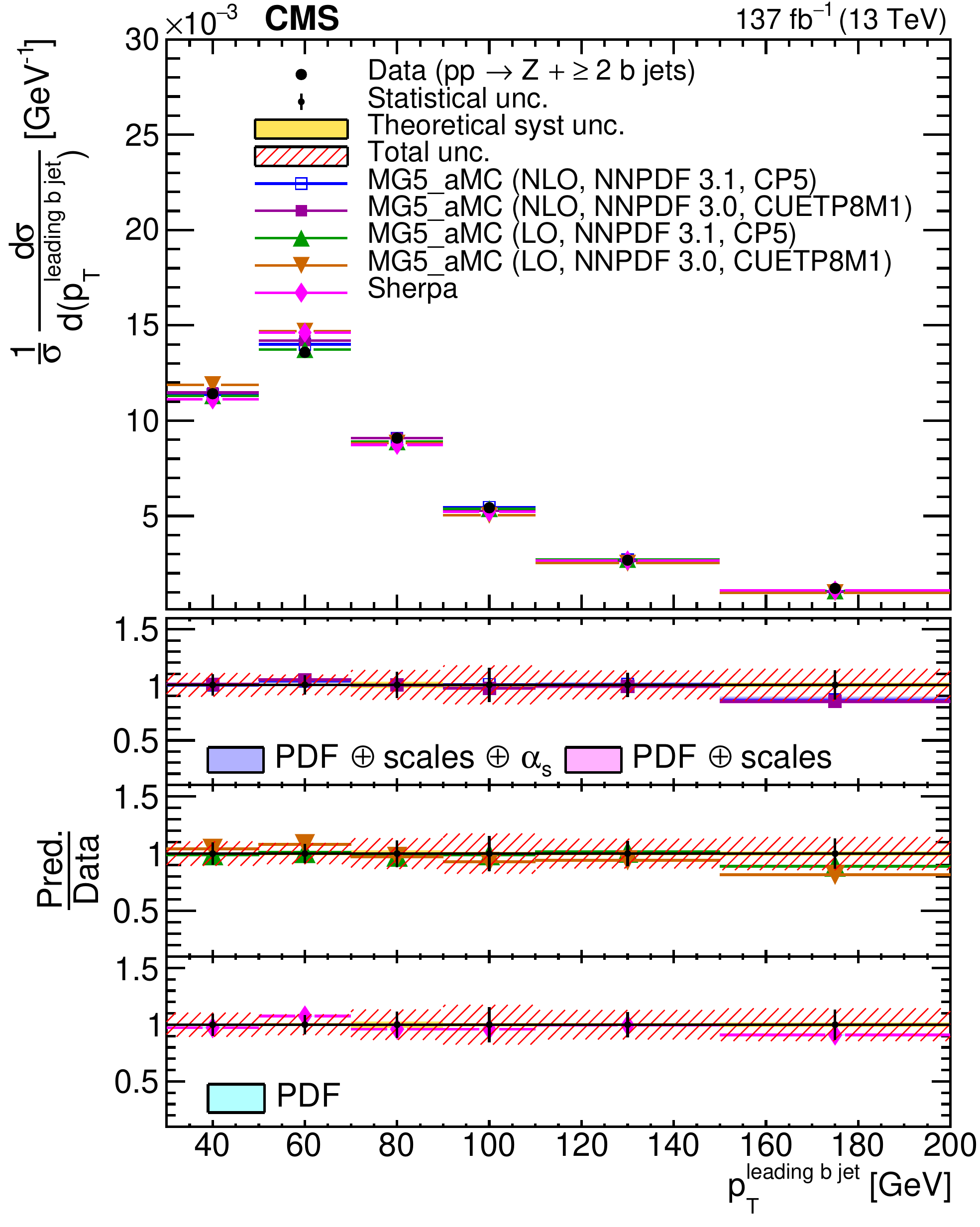}}
\caption{(\cmsLeft) Differential cross section and the (\cmsRight) normalized differential cross section distributions as functions of the leading \PQb jet transverse momentum for the \zgttwob events.}
\label{fig:diff_xsec_pt_b1}
\end{figure*}

\begin{figure*}[htb]
\centering
{\includegraphics[width=0.44\textwidth]{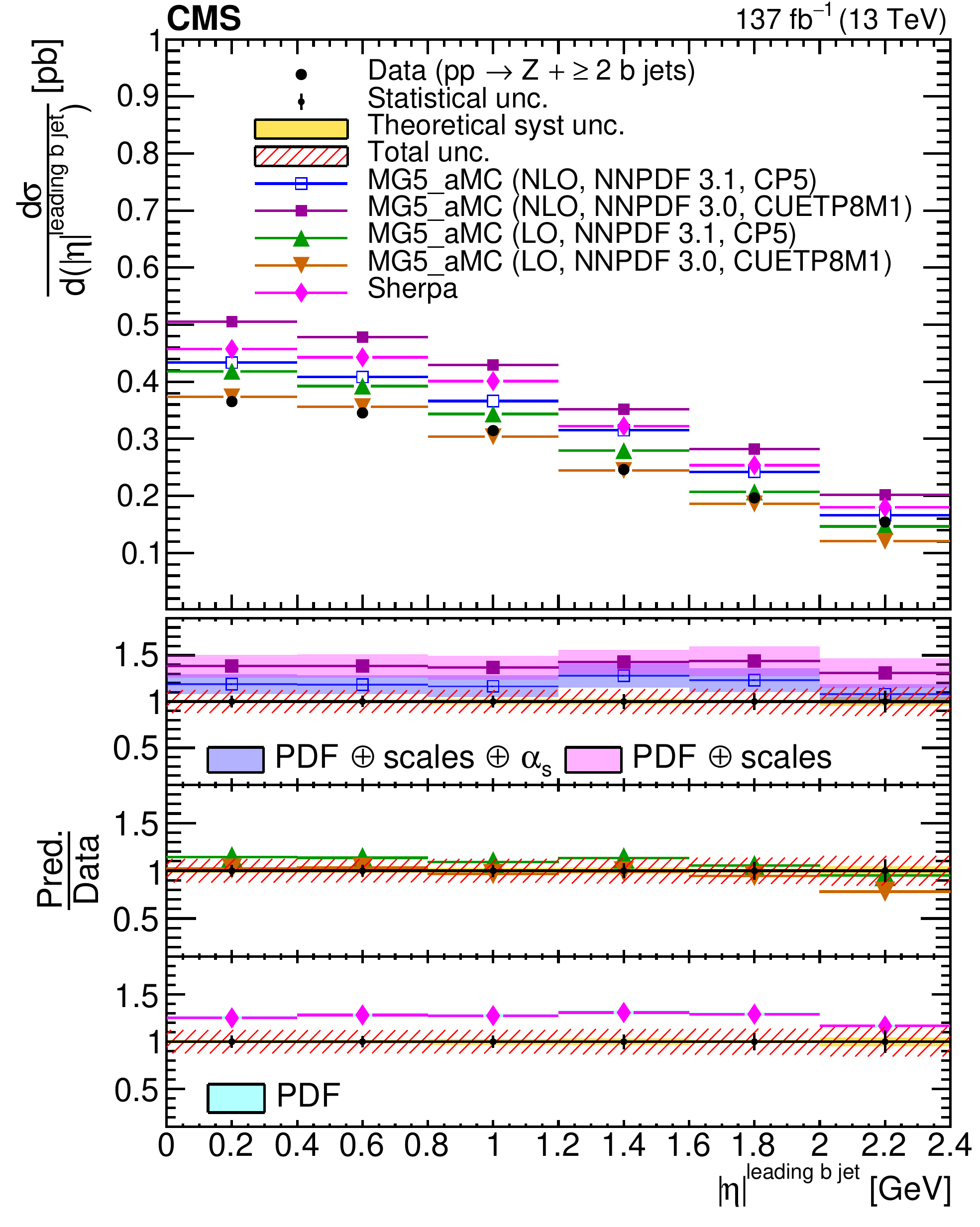}}
{\includegraphics[width=0.44\textwidth]{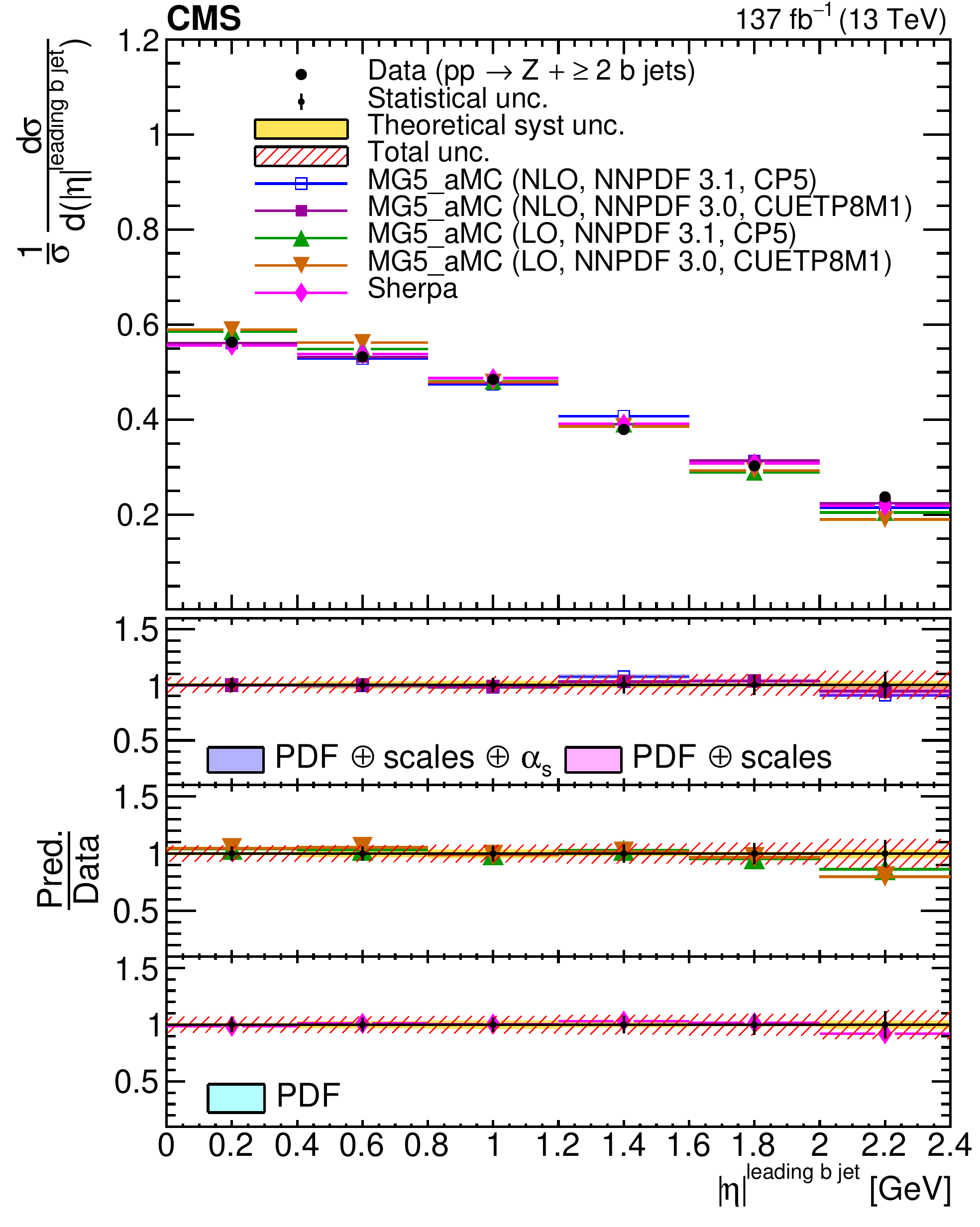}}
\caption{(\cmsLeft) Differential cross section and the (\cmsRight) normalized differential cross section distributions as functions of the leading \PQb jet absolute pseudorapidity for the \zgttwob events.}
\label{fig:diff_xsec_abseta_b1}
\end{figure*}

\begin{figure*}[htb]
\centering
{\includegraphics[width=0.44\textwidth]{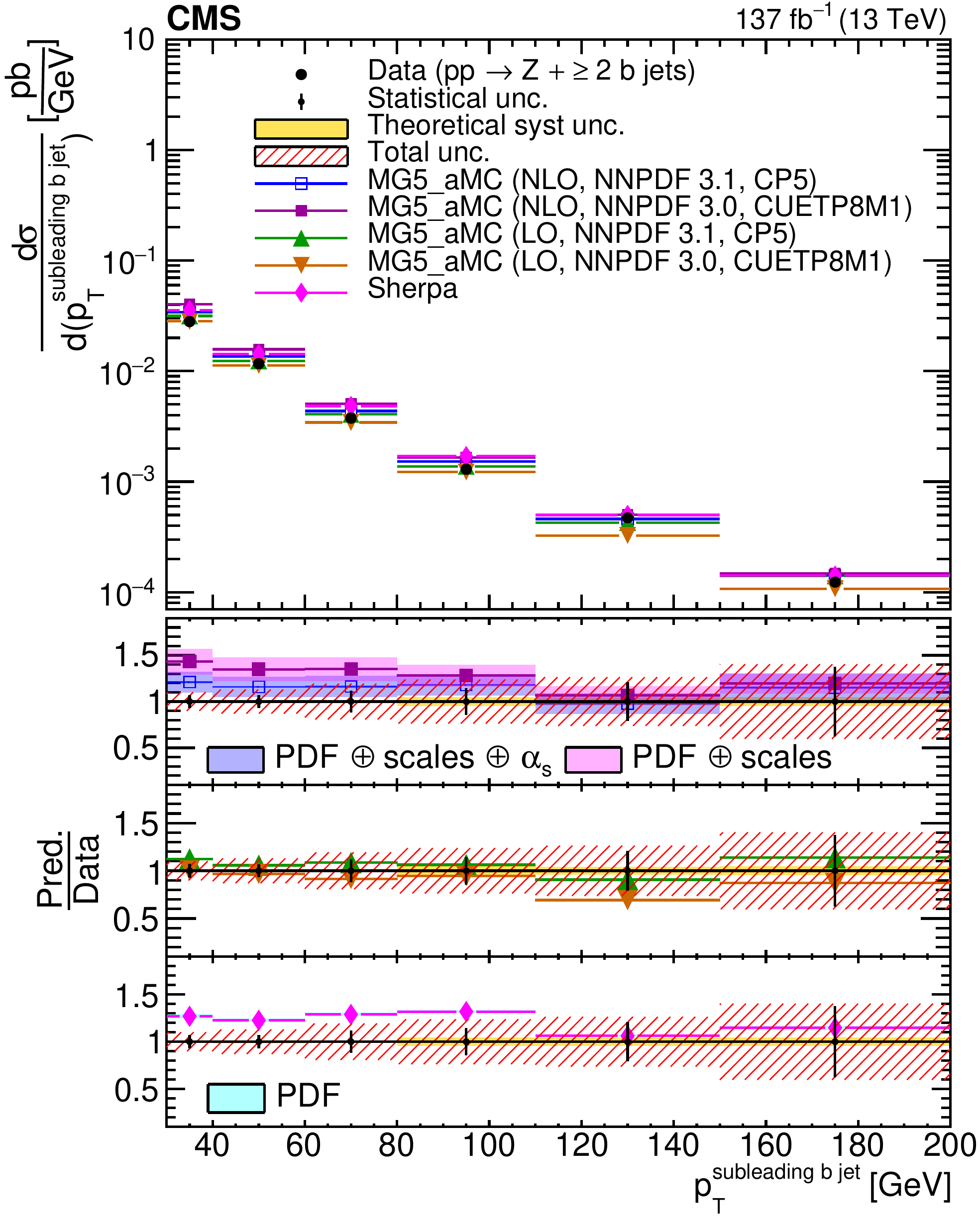}}
{\includegraphics[width=0.44\textwidth]{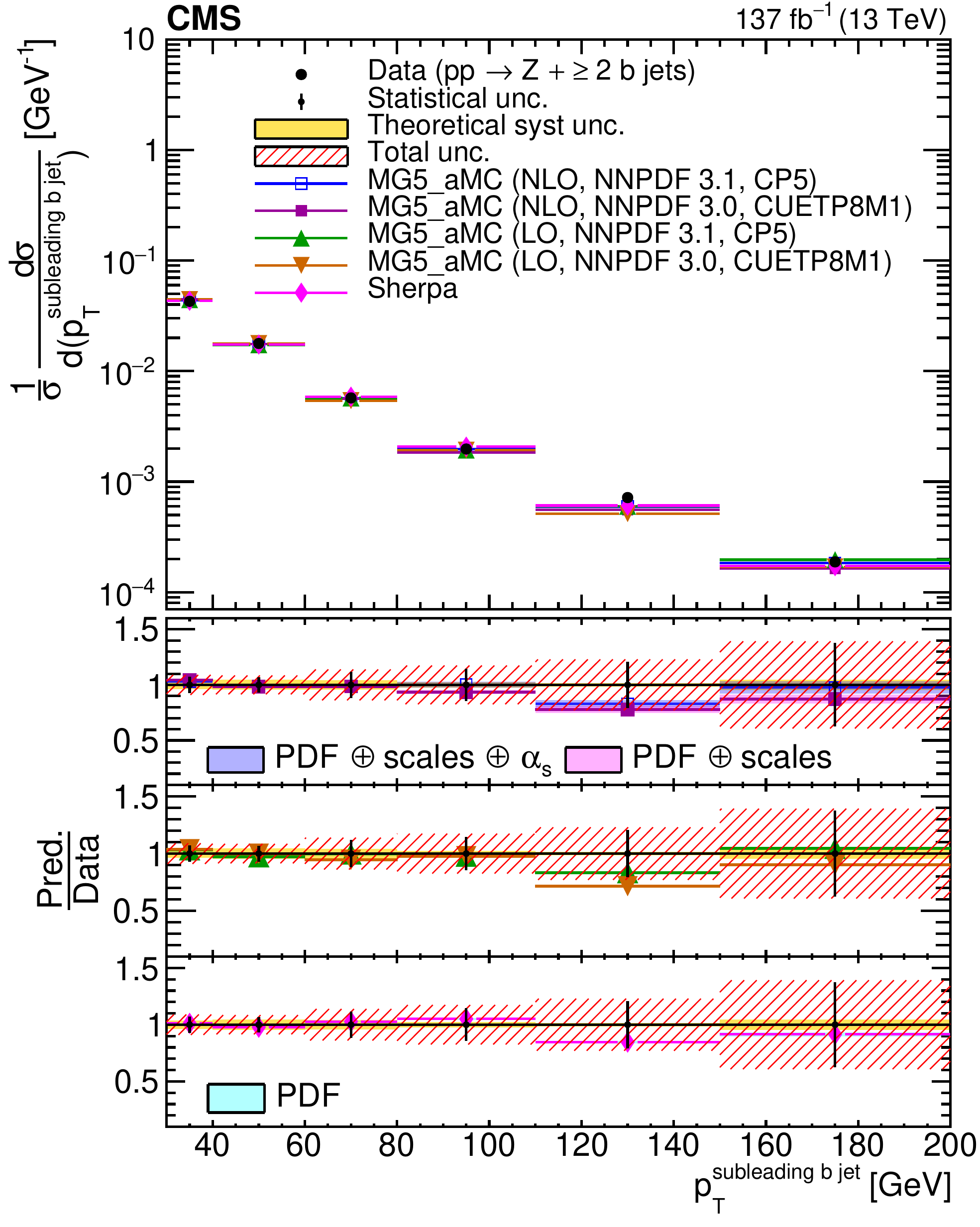}}
\caption{(\cmsLeft) Differential cross section and the (\cmsRight) normalized differential cross section distributions as functions of the subleading \PQb jet transverse momentum for the \zgttwob events.}
\label{fig:diff_xsec_pt_b2}
\end{figure*}

\begin{figure*}[htb]
\centering
{\includegraphics[width=0.44\textwidth]{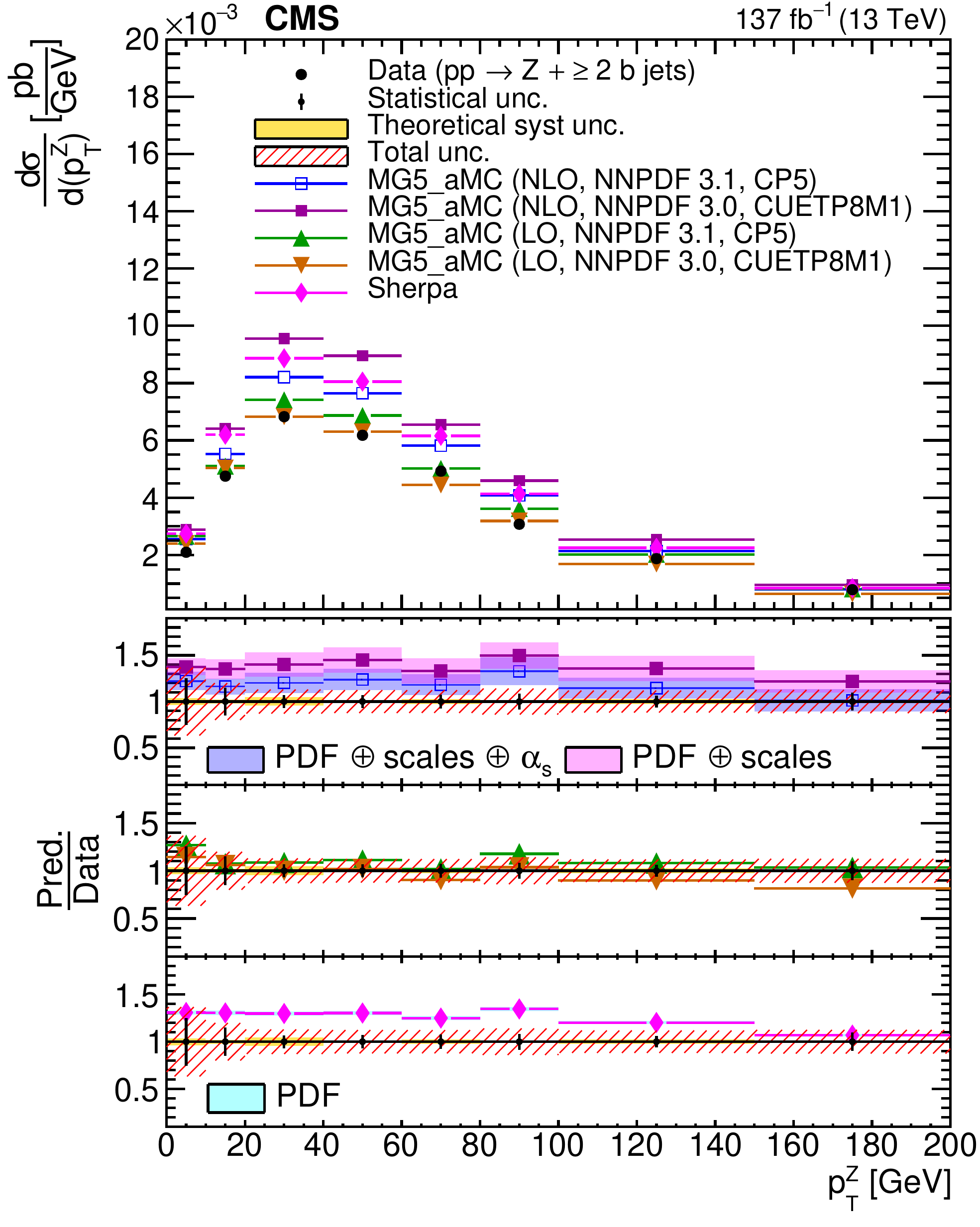}}
{\includegraphics[width=0.44\textwidth]{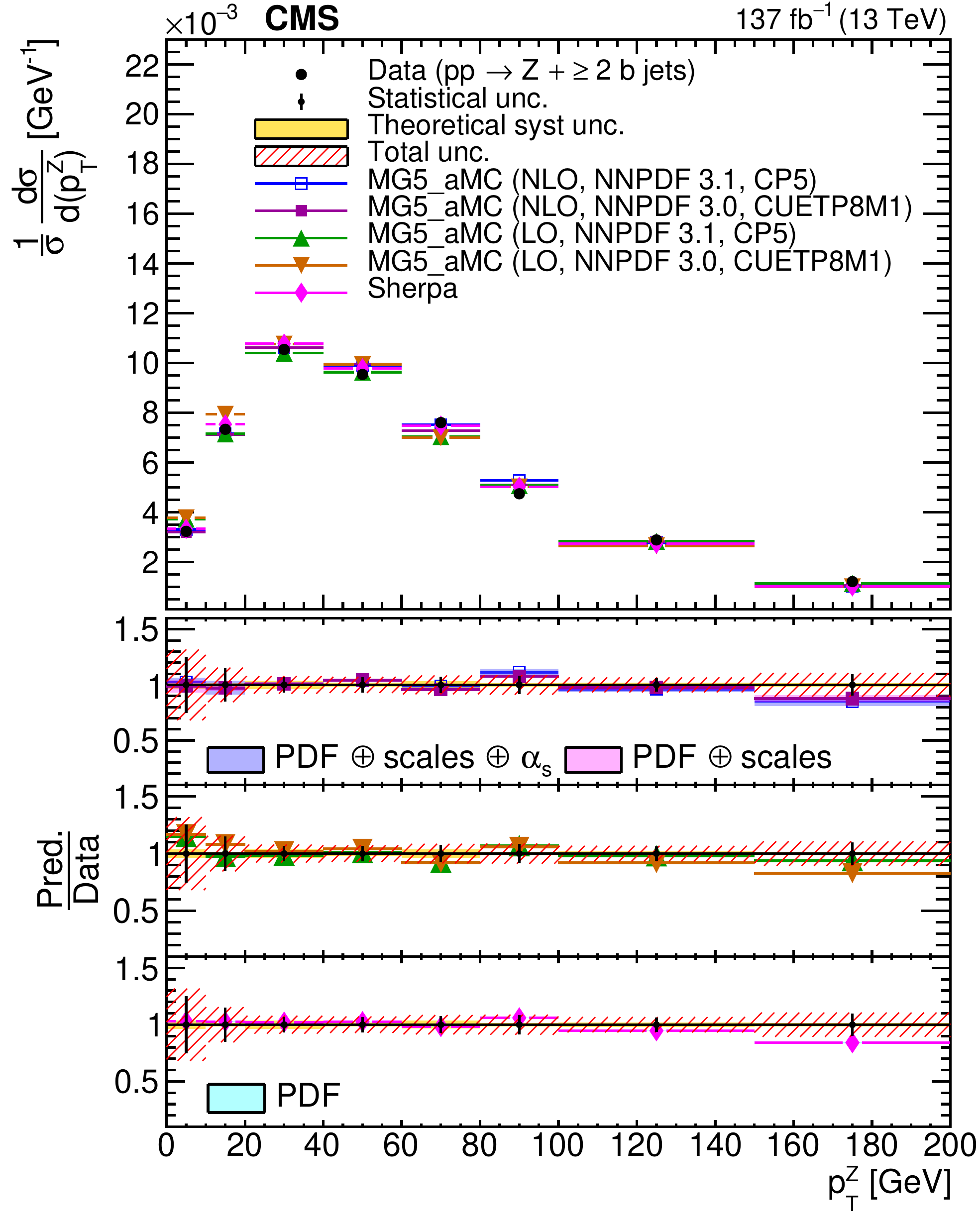}}
\caption{(\cmsLeft) Differential cross section and the (\cmsRight) normalized differential cross section distributions as functions of the \PZ boson transverse momentum for the \zgttwob events.}
\label{fig:diff_xsec_pt_Z}
\end{figure*}

\begin{figure*}[htb]
\centering
{\includegraphics[width=0.44\textwidth]{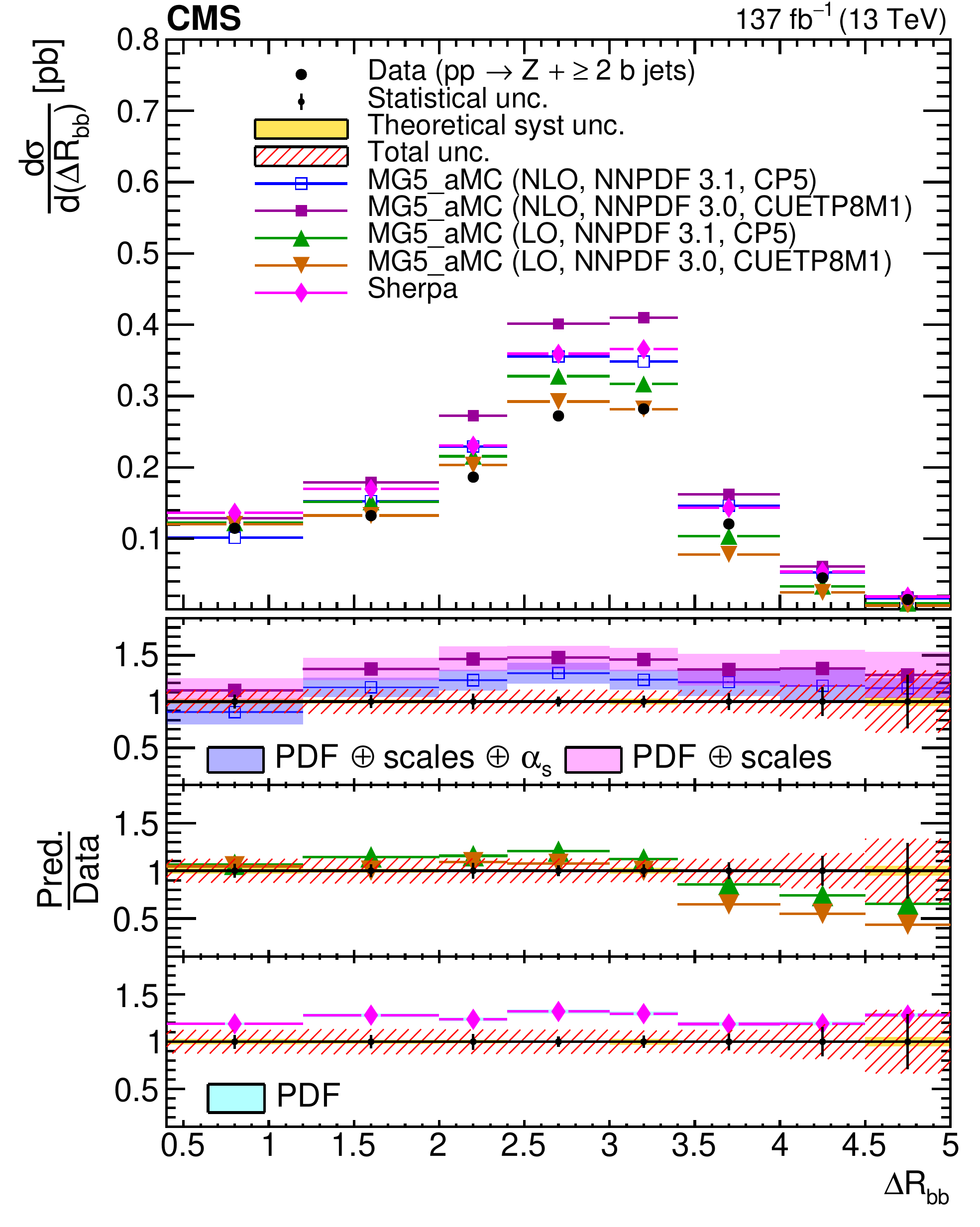}}
{\includegraphics[width=0.44\textwidth]{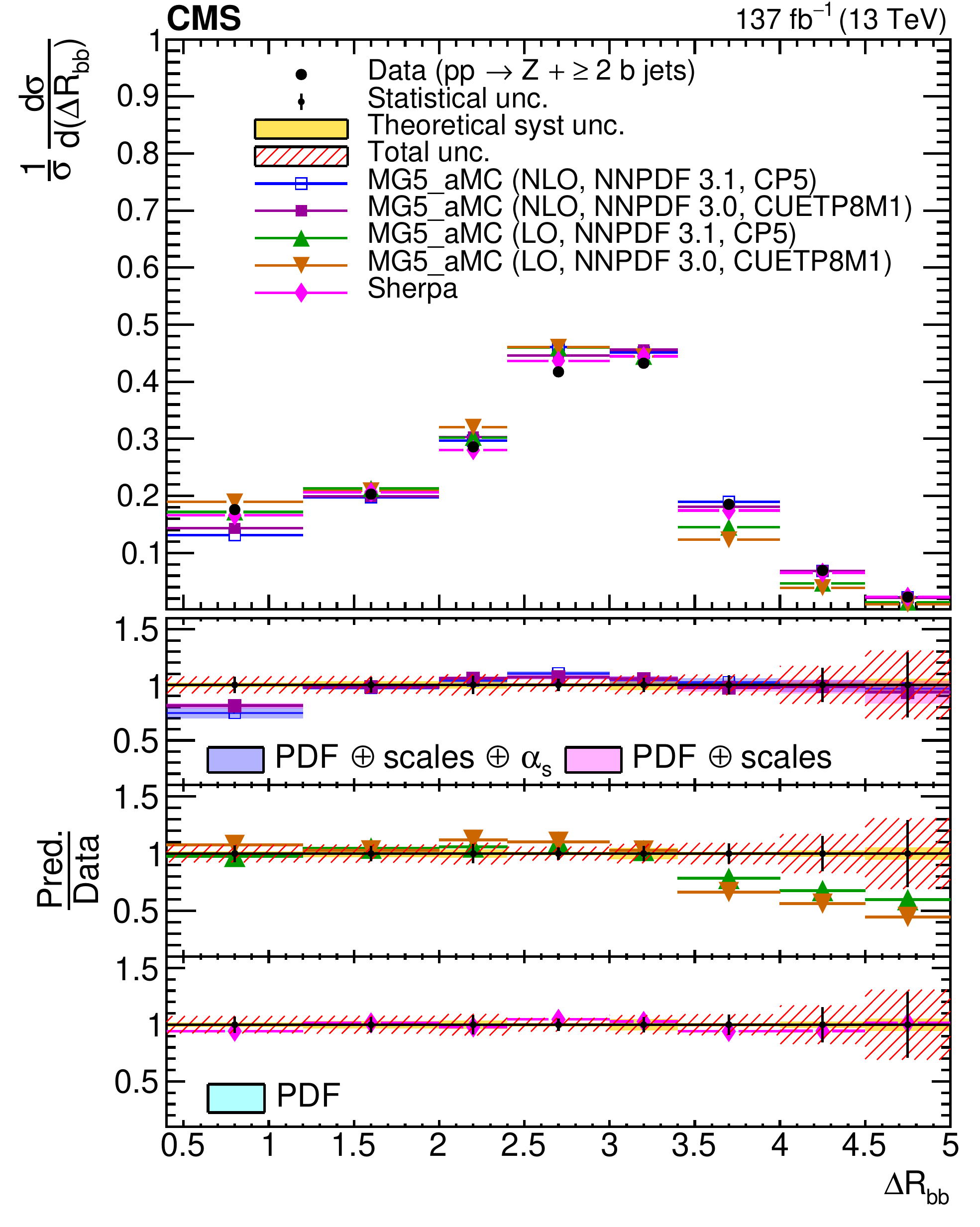}}
\caption{(\cmsLeft) Differential cross section and the (\cmsRight) normalized differential cross section distributions as functions of the angular separation between two \PQb jets, $\Delta R_{\mathrm{\PQb\PQb}}$ for the \zgttwob events.}
\label{fig:diff_xsec_dR_2b}
\end{figure*}

\begin{figure*}[htb]
\centering
{\includegraphics[width=0.44\textwidth]{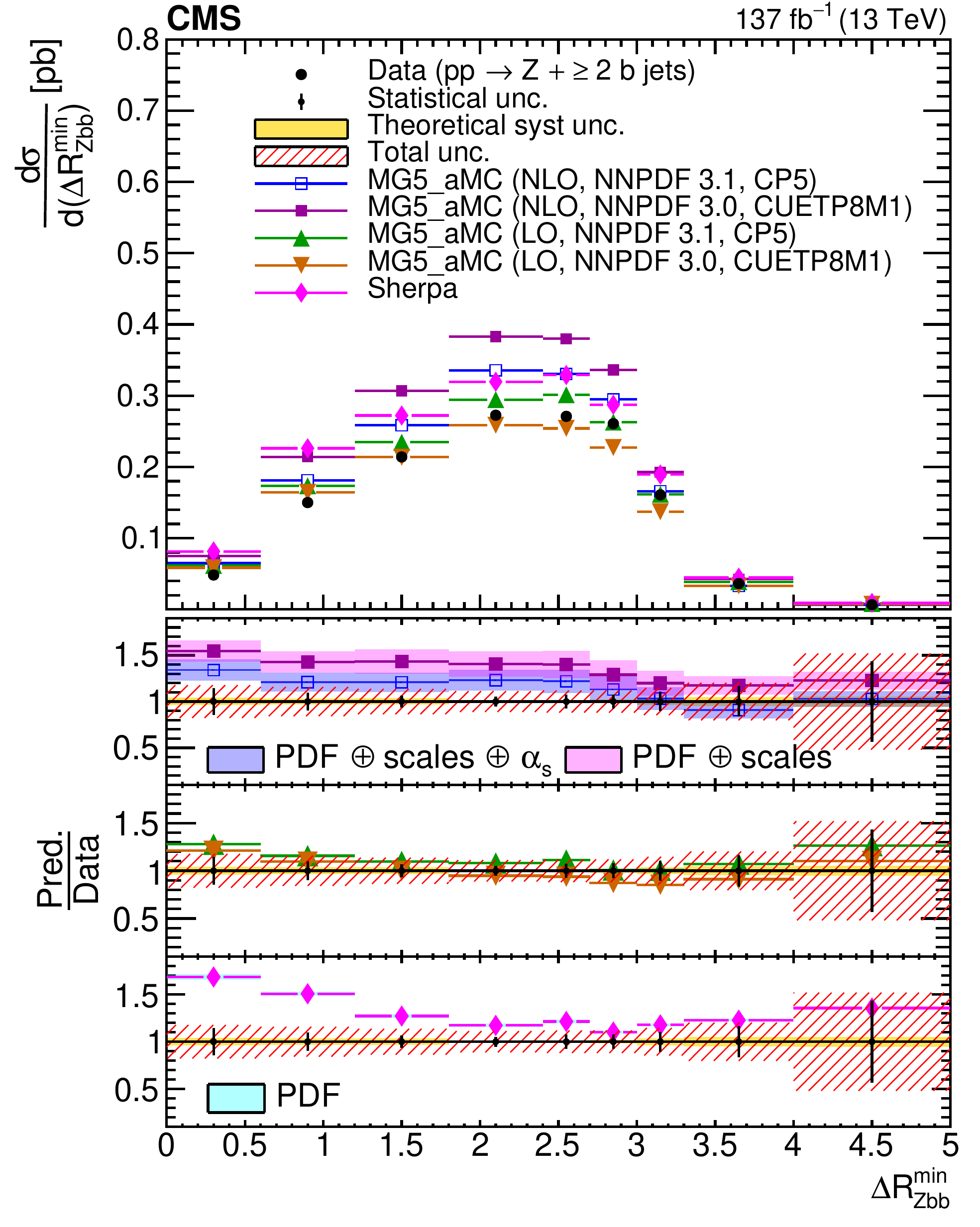}}
{\includegraphics[width=0.44\textwidth]{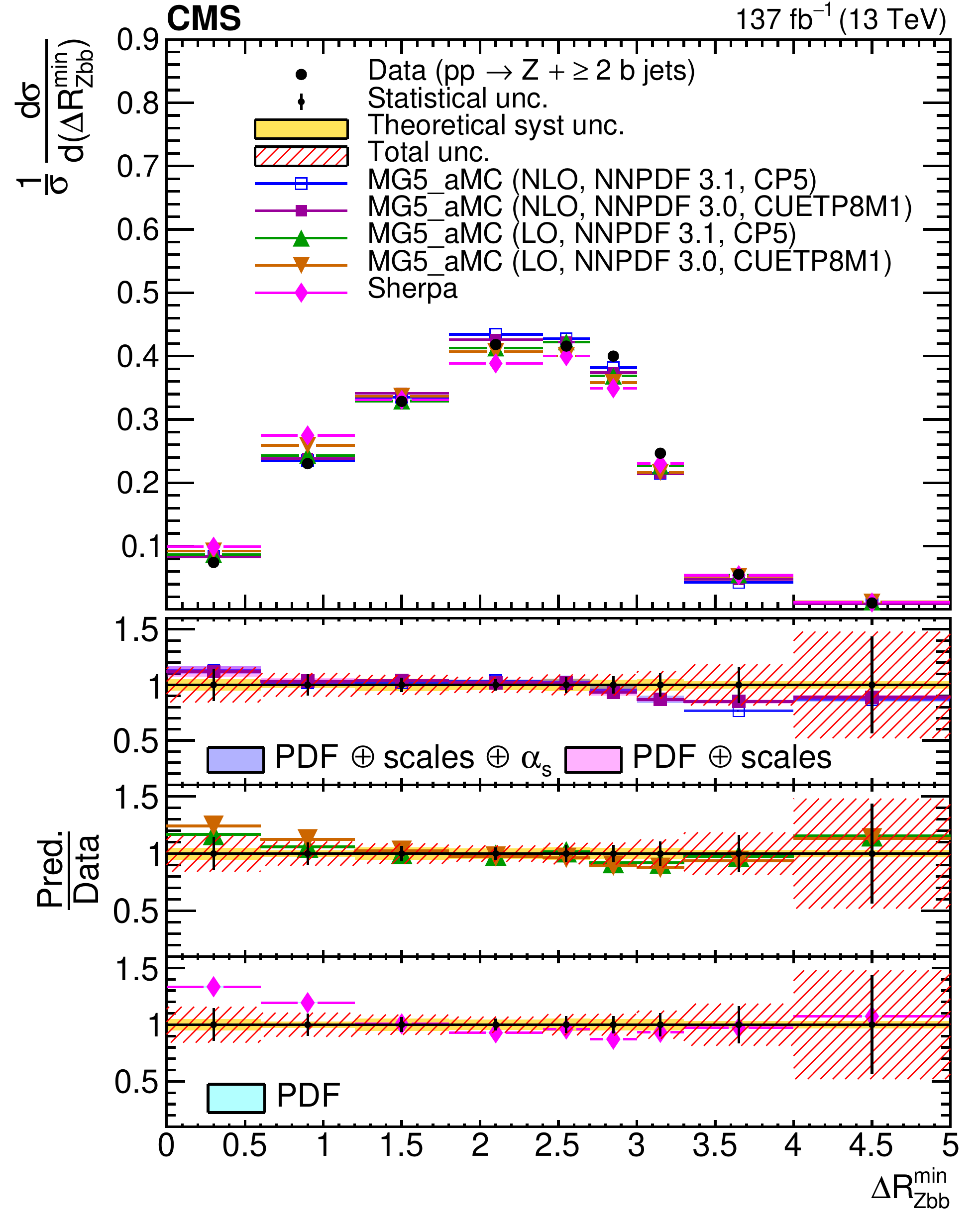}}
\caption{(\cmsLeft) Differential cross section and the (\cmsRight) normalized differential cross section distributions as functions of the minimum angular separation between the \PZ boson and two \PQb jets, $\Delta R^{\text{min}}_{\mathrm{Zbb}}$ for the \zgttwob events.}
\label{fig:diff_xsec_dRmin_Z2b}
\end{figure*}

\begin{figure*}[htb]
\centering
{\includegraphics[width=0.44\textwidth]{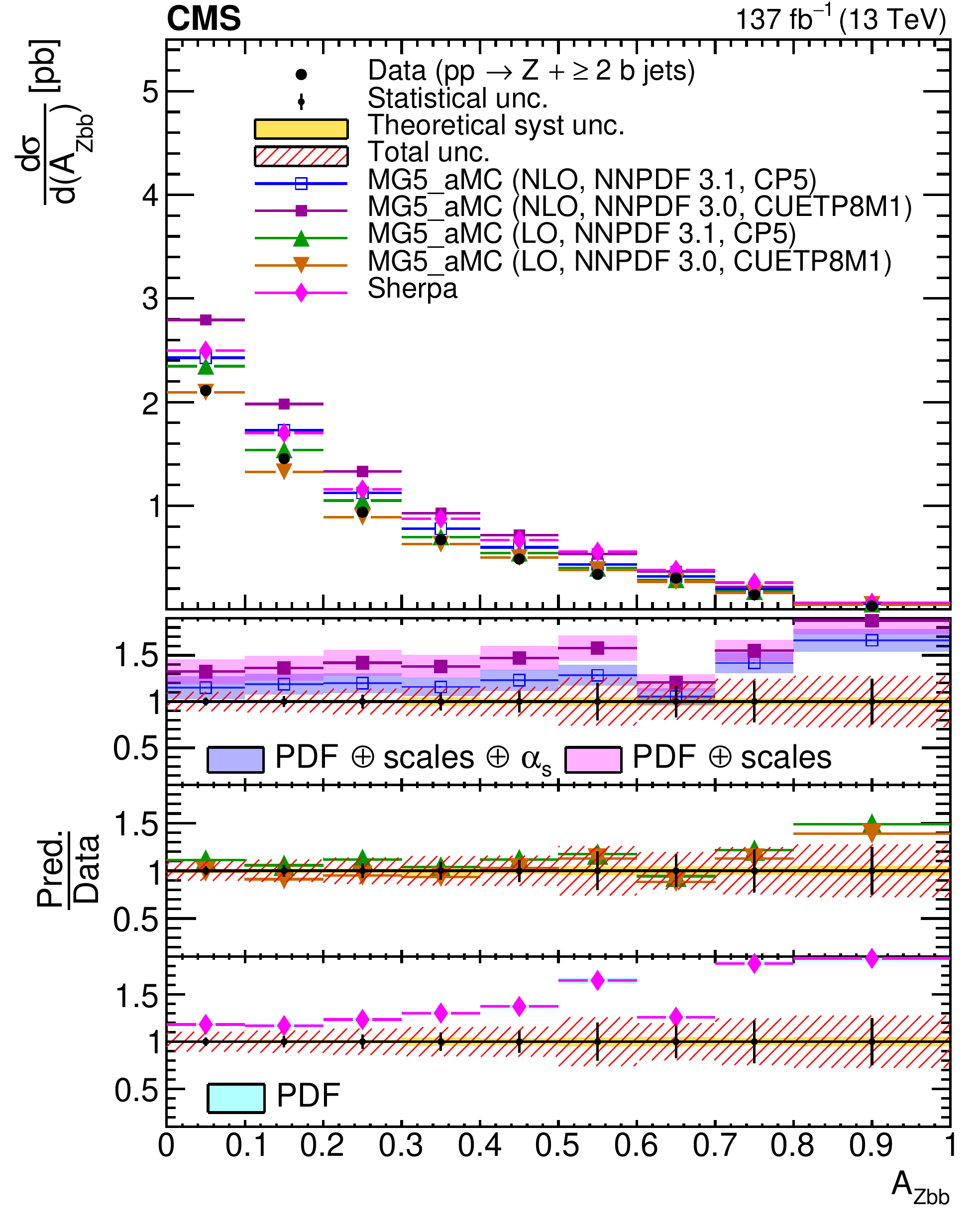}}
{\includegraphics[width=0.44\textwidth]{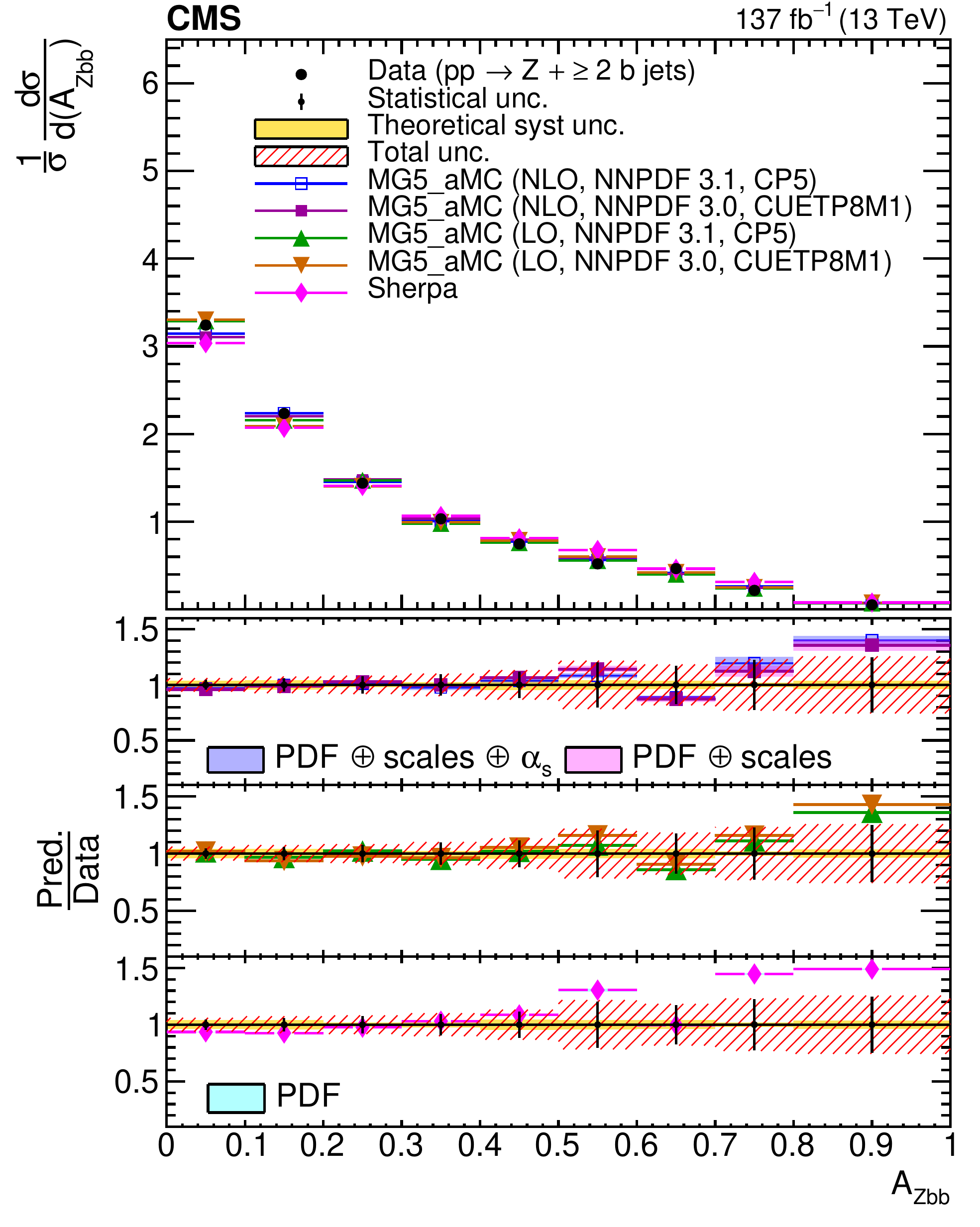}}
\caption{(\cmsLeft) Differential cross section and the (\cmsRight) normalized differential cross section distributions as functions of the asymmetry of the \zgttwob system, $A_{\mathrm{Zbb}}$.}
\label{fig:diff_xsec_A_Z2b}
\end{figure*}

\begin{figure*}[htb]
\centering
{\includegraphics[width=0.44\textwidth]{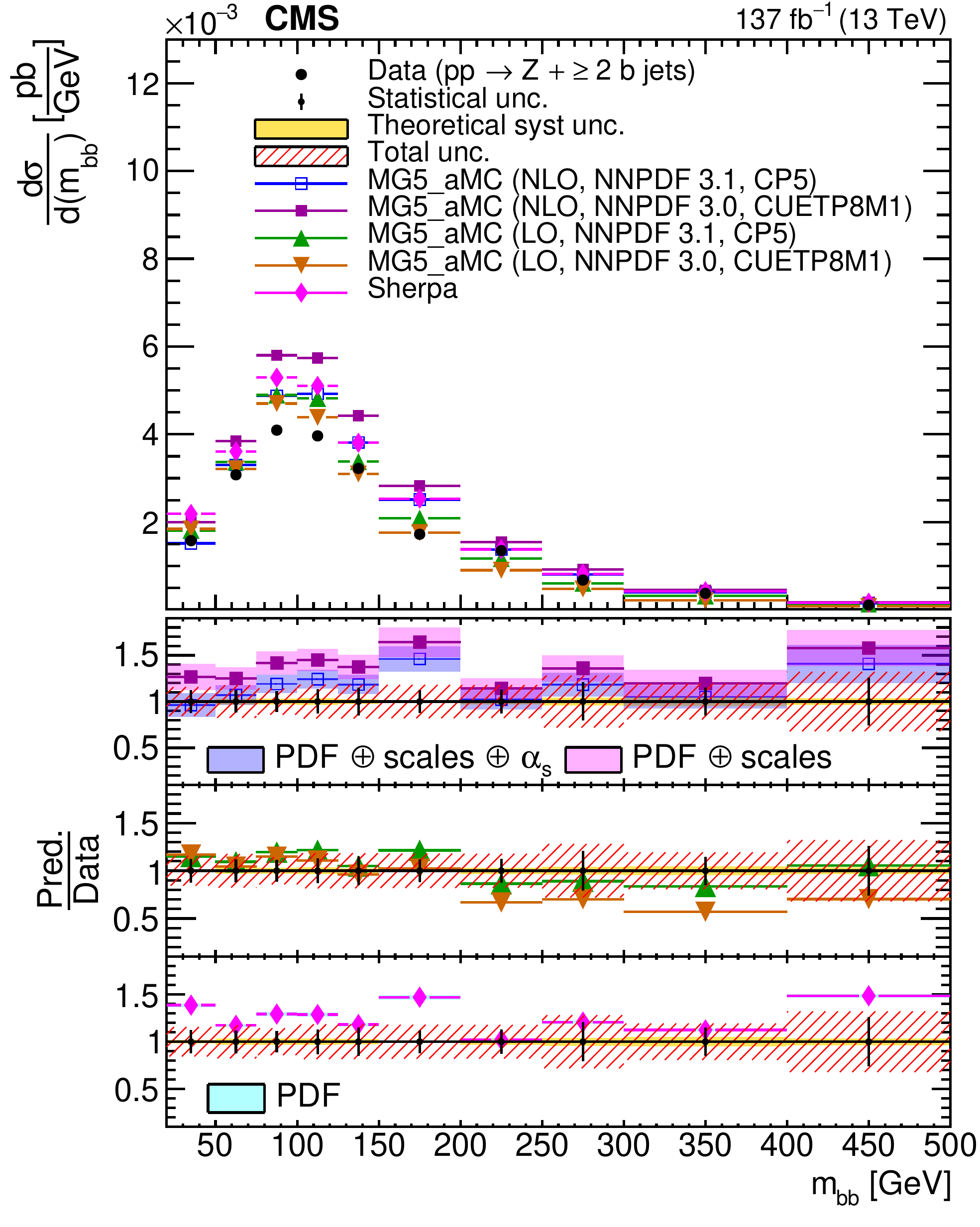}}
{\includegraphics[width=0.44\textwidth]{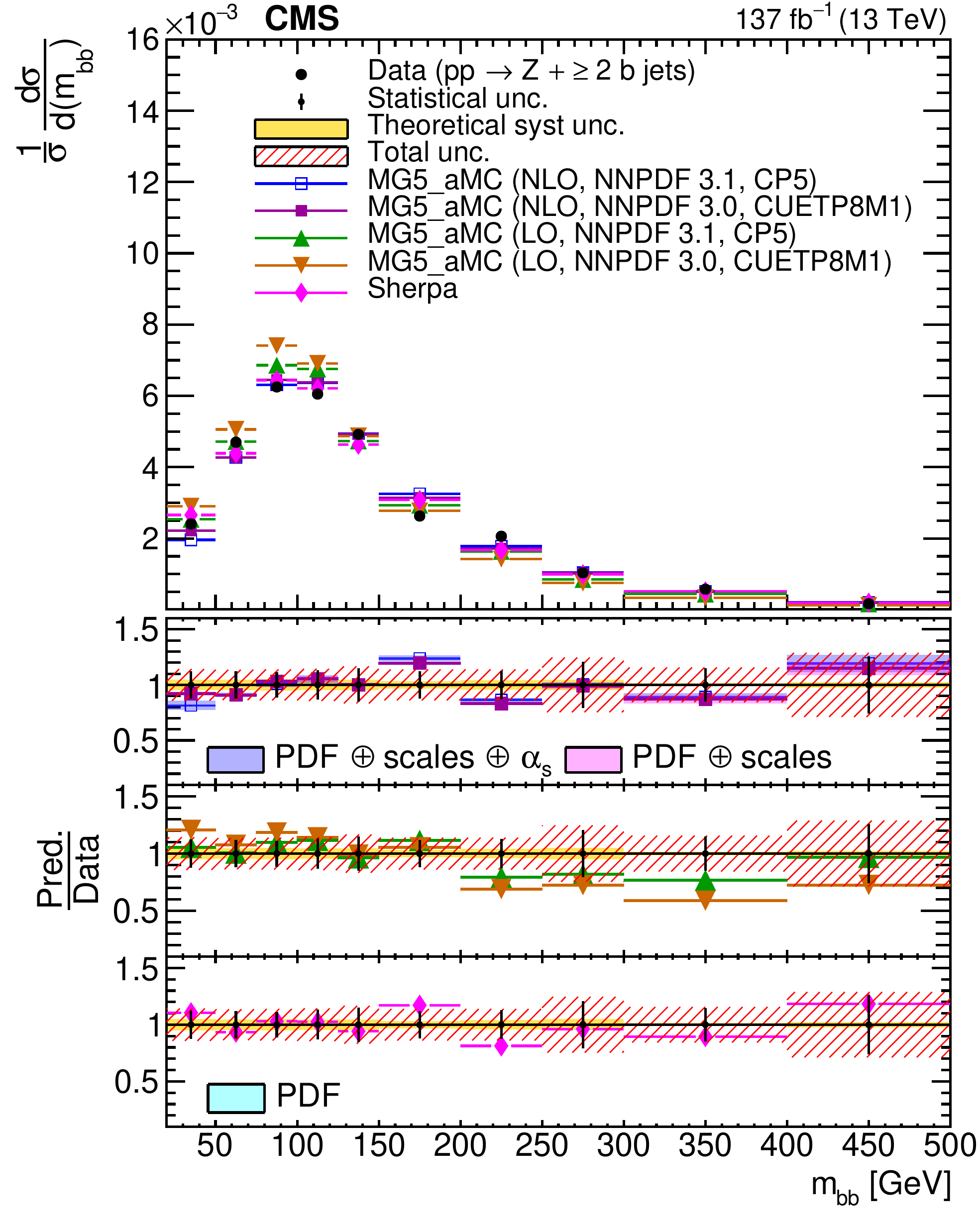}}
\caption{(\cmsLeft) Differential cross section and the (\cmsRight) normalized differential cross section distributions as functions of the invariant mass of the two \PQb jets.}
\label{fig:diff_xsec_m_2b}
\end{figure*}

\begin{figure*}[htb]
\centering
{\includegraphics[width=0.44\textwidth]{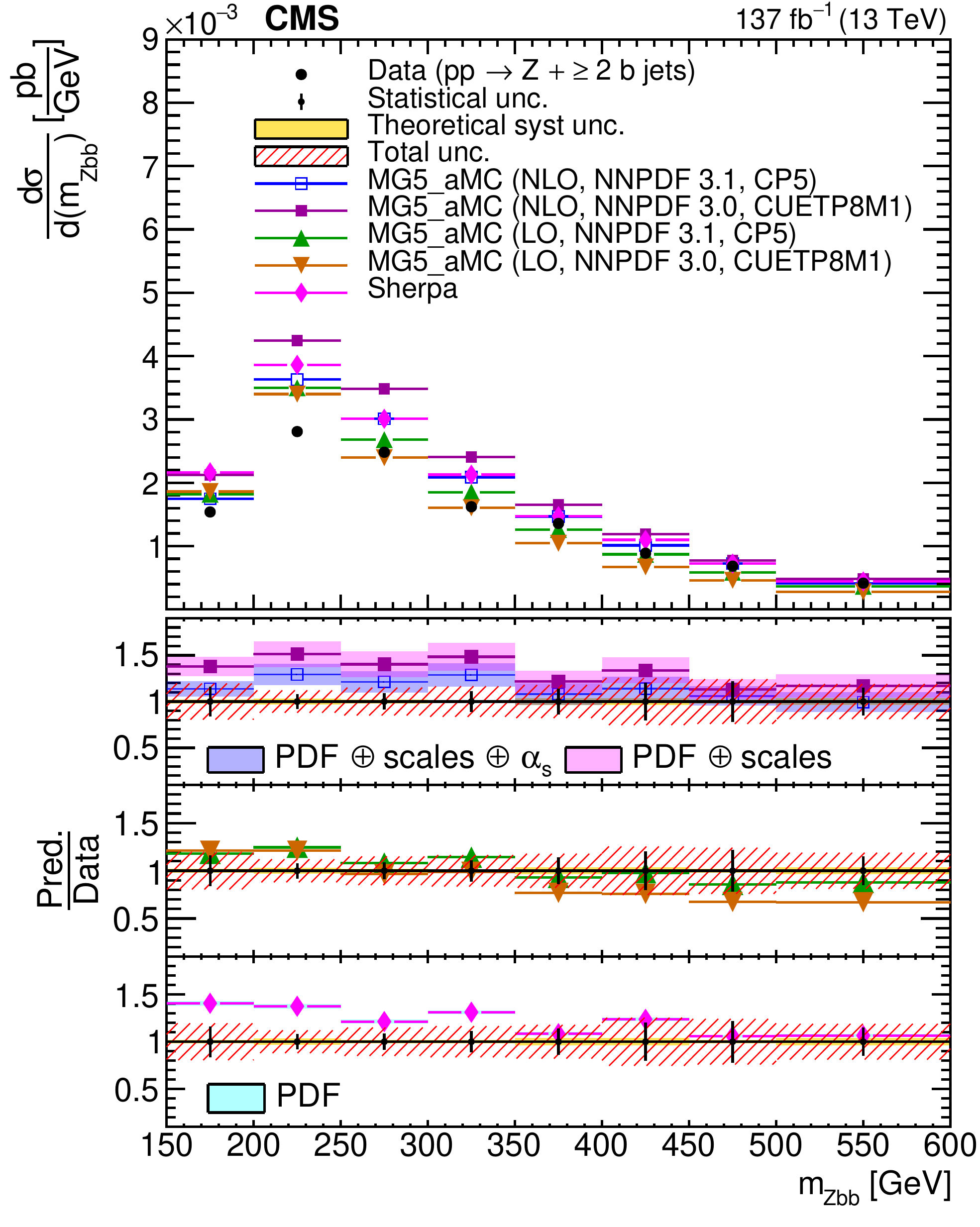}}
{\includegraphics[width=0.44\textwidth]{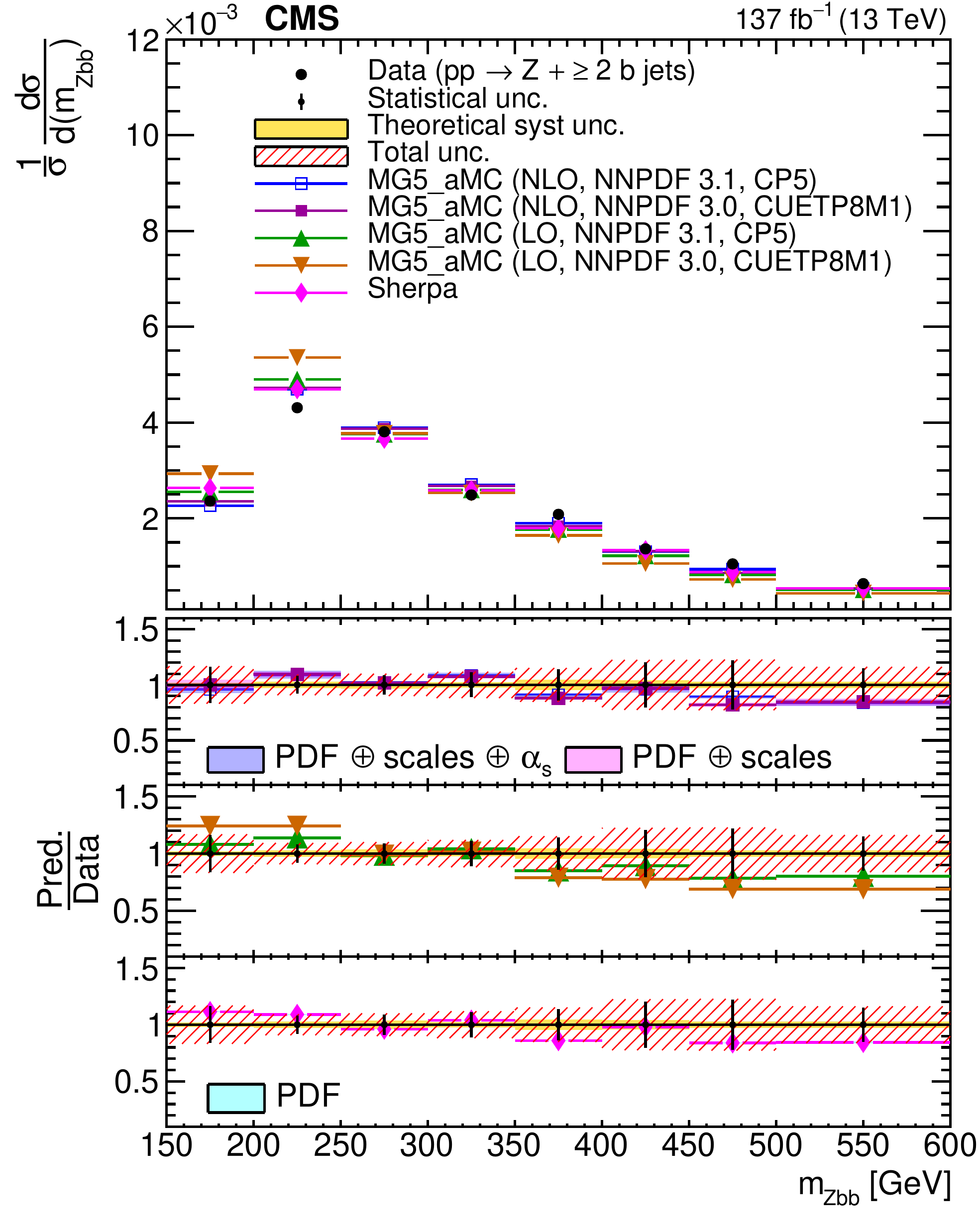}}
\caption{(\cmsLeft) Differential cross section and the (\cmsRight) normalized differential cross section distributions as functions of the invariant mass of the \PZ boson and two \PQb jets.}
\label{fig:diff_xsec_m_Z2b}
\end{figure*}

The measured differential cross section distributions and the corresponding ones normalized by the integrated fiducial cross sections compared with different MC predictions in the combined channel are shown in Figs.~\ref{fig:cross_dilepton_pt}--\ref{fig:diff_xsec_m_Z2b}. In general, the differential cross section distributions predicted by \SHERPA tend to overestimate data by 20--30\%, depending on the kinematic variables.

The differential and normalized differential cross section distributions as functions of the \ZpT in the selected \zgtoneb events are shown in Fig.~\ref{fig:cross_dilepton_pt}. The shapes of these distributions are described best by \MGfive (LO, NNPDF 3.1, CP5), while \MGfive (NLO) and \SHERPA predictions vary up to 30\% depending on \ZpT. The \bjetpT and \bjetabseta differential and normalized differential cross section distributions are shown in Figs.~\ref{fig:cross_dileptonbjetpt_pt} and~\ref{fig:cross_dileptonbjet_eta}, respectively. The shapes of distributions are well described by all simulations, except for the \MGfive (LO, NNPDF 3.1, CP5) \bjetpT spectrum, which deviates up to 25\% in the higher \pt region. The differential and normalized differential cross section distributions of \deltaPhi, \deltaY, and \deltaR are shown in Figs.~\ref{fig:cross_dilepton_delphi}--\ref{fig:cross_dilepton_delR}. The shapes of these distributions are best described by the \SHERPA predictions. The \MGfive (LO) predictions show the largest deviation from data in the high \deltaY and \deltaR regions, which is significantly improved with the \MGfive (NLO) prediction. In summary, \SHERPA simulations provide a good description of different kinematic observables except for the higher \ZpT region for the normalized differential distributions. The \MGfive LO and NLO simulations provide varying levels of agreement for the observables.


For the \zgttwob final states, differential and the normalized differential cross sections as functions of the leading \PQb jet \pt and $\abs{\eta}$, subleading \PQb jet \pt, and \PZ boson transverse momentum are shown in Figs.~\ref{fig:diff_xsec_pt_b1}--\ref{fig:diff_xsec_pt_Z}. The shapes of distributions in data are well-described by predictions from \MGfive (LO and NLO; with both PDFs) and \SHERPA, except for a couple of bins in the high-\pt and high-$\abs{\eta}$ regions. In the differential cross sections \MGfive (NLO, NNPDF 3.1, CP5), \MGfive (NLO, NNPDF 3.0, CUETP8M1), and \SHERPA overestimate data by 20--30\%, 30--50\%, and 20--30\%, respectively, whereas the \MGfive (LO) predictions are in good agreement with data. The angular correlations and asymmetry distributions of the \zgttwob system are presented in Figs.~\ref{fig:diff_xsec_dR_2b}--\ref{fig:diff_xsec_A_Z2b}. Although \MGfive (NLO) predictions are consistent with data in the normalized differential cross section distributions, except for the $\smash{\Delta R_{\PQb\PQb}}$, $A_{\PZ\PQb\PQb}$ regions of 0.5--1.5 and 0.8--1, respectively, they are generally higher than the measured differential cross sections by 20--30\% and 30--50\% for \MGfive (NLO, NNPDF 3.1, CP5) and \MGfive (NLO, NNPDF 3.0, CUETP8M1), respectively. The \MGfive (LO) predictions show large deviations from data in the $\smash{\Delta R_{\PQb\PQb}}$ distributions above 3.4, as can be seen in Fig.~\ref{fig:diff_xsec_dR_2b}. In the $\Delta R^{\text{min}}_{\PZ\PQb\PQb}$ distribution, Fig.~\ref{fig:diff_xsec_dRmin_Z2b}, the \MGfive (LO) predictions are consistent with data, although they are higher than data in the last bin of the $A_{\PZ\PQb\PQb}$ distribution (0.8--1) in Fig.~\ref{fig:diff_xsec_A_Z2b}, similar to the \MGfive (NLO, NNPDF 3.1, CP5) case. The \MGfive (NLO, NNPDF 3.0, CUETP8M1) simulation overestimates data in the differential $\Delta R^{\text{min}}_{\PZ\PQb\PQb}$ distribution. The \SHERPA predictions describe well the shape of the measured $\smash{\Delta R_{\PQb\PQb}}$ differential cross sections but they significantly overestimate data in the low ($<$1.2) and high ($>$0.5) regions of the $\Delta R^{\text{min}}_{\PZ\PQb\PQb}$ and $A_{\PZ\PQb\PQb}$ distributions, respectively. Figure~\ref{fig:diff_xsec_m_2b} shows invariant mass for two \PQb jets events, and Fig.~\ref{fig:diff_xsec_m_Z2b} shows the invariant mass for the \zgttwob events. The shapes of these distributions are described better by the \MGfive (NLO) and \SHERPA predictions than the \MGfive (LO) ones. As for the rates, the data are described best by the \MGfive (NLO, NNPDF 3.1, CP5) predictions.

\begin{figure*}[htb]
\centering
{\includegraphics[width=0.44\textwidth]{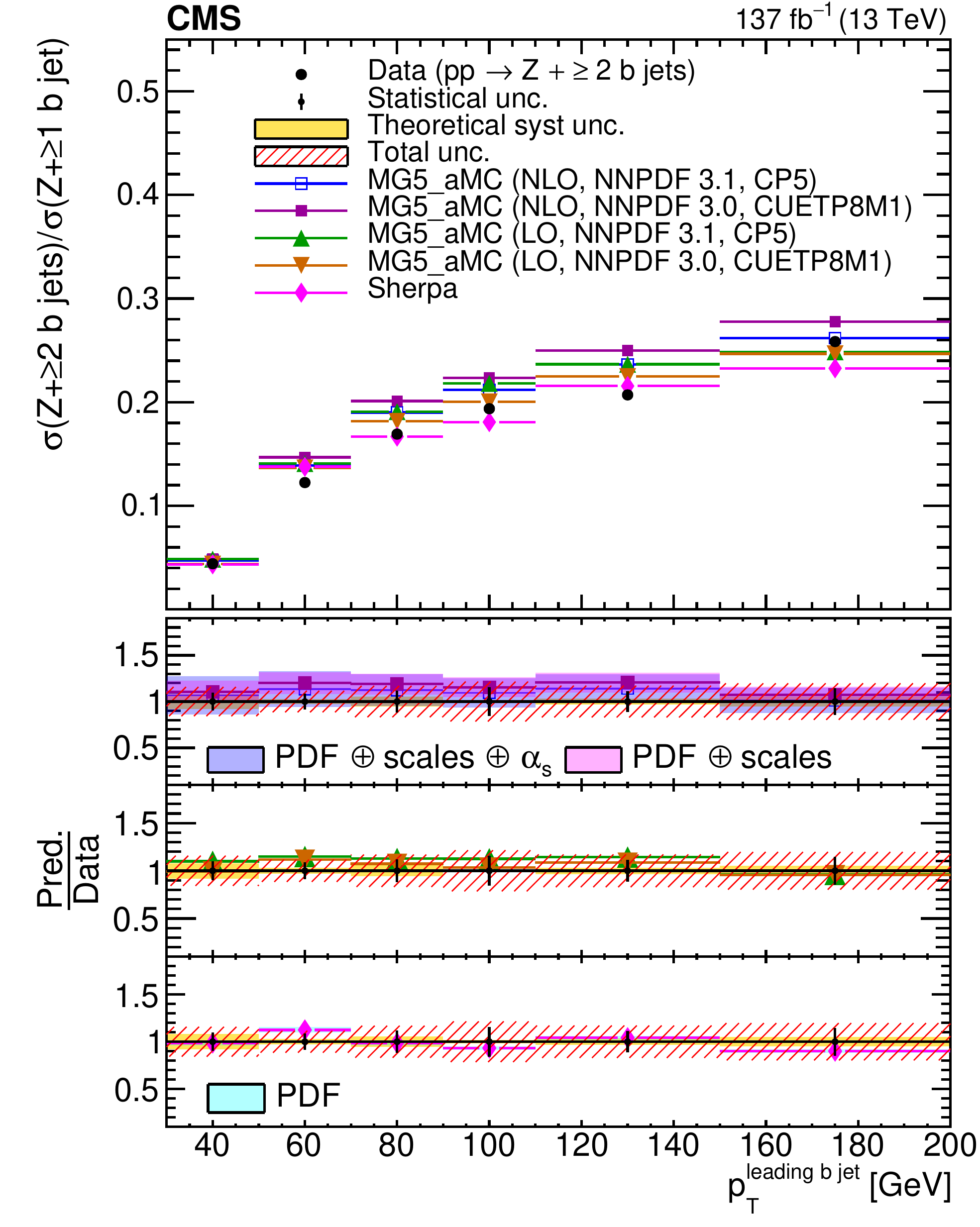}}
{\includegraphics[width=0.44\textwidth]{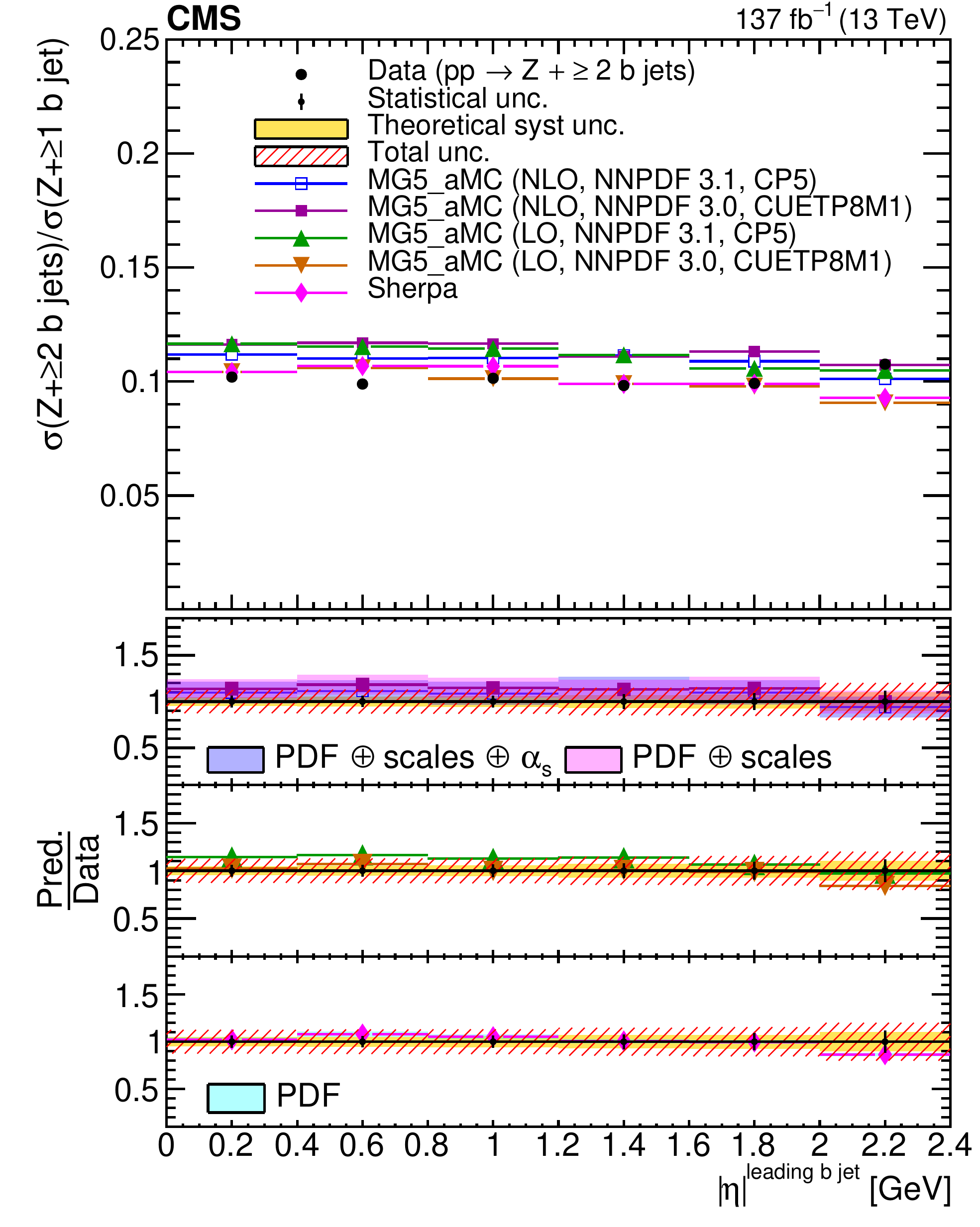}}
\caption{Distributions of the cross section ratios, $\sigma$(\zgttwob)/$\sigma$(\zgtoneb), as functions of the (\cmsLeft) leading \PQb jet transverse momentum and (\cmsRight) absolute pseudorapidity.}
\label{fig:xsec_rat}
\end{figure*}

The $\sigma$(\zgttwob)/$\sigma$(\zgtoneb) cross section ratio distributions as functions of the leading \PQb jet transverse momentum and absolute pseudorapidity are shown in Fig.~\ref{fig:xsec_rat}. The ratio gradually increases (from 0.05 to 0.25) with the leading \PQb jet \pt (ranging from 30 to 200\GeV), but is nearly independent of the pseudorapidity of the leading \PQb jet. The increase in the ratio is due to a kinematic effect; at larger \pt of the leading jet, the kinematic acceptance for the subleading jet would increase. Predictions from \MGfive (LO), \MGfive (NLO), and \SHERPA describe the measured ratios within uncertainties.


\clearpage
\section{Summary}
A measurement of fiducial cross sections of the \zgtoneb and \zgttwob processes, along with the differential and normalized differential cross section distributions of different kinematic observables, is performed using proton-proton collisions data at $\sqrt{s}=$ 13\TeV collected by the CMS experiment at the CERN LHC. This is the first measurement of these processes based on data collected during the 2016--2018 LHC running period. The fiducial cross sections are measured to be $6.52\pm 0.04\stat\pm 0.40\syst\pm 0.14\thy$\unit{pb} for the \zgtoneb and $0.65\pm 0.03\stat\pm 0.07\syst\pm 0.02 \thy$\unit{pb} for the \zgttwob, which are better described by the \MGfive leading order (LO) simulation but overestimated by \MGfive next-to-LO (NLO) and \SHERPA predictions. Since all predictions are normalized to the inclusive \zjs next-to-NLO cross section, differences between \MGfive (NLO) and \MGfive (LO) results could be attributable to variations in shapes of observables and settings (parton distribution functions, Monte Carlo tunes, matching schemes) used in those simulations. The \SHERPA simulation overestimates the measured integrated cross section; however, it provides a good description of the shapes of various kinematic observables. The \MGfive (LO) and \MGfive (NLO) generators interfaced with {\PYTHIA} describe the fiducial cross section better but do not completely describe the shapes of the kinematic observables. Present measurements can be used as an input for the further optimization of the simulation parameters. The measured value of the cross section ratio of the \zgttwob to \zgtoneb is $0.100\pm 0.005\stat\pm 0.007\syst\pm 0.003\thy$, which is well described by the \MGfive (LO, NNPDF 3.0, CUETP8M1) and \SHERPA calculations but overestimated by \MGfive (NLO) prediction.

\begin{acknowledgments}
  We congratulate our colleagues in the CERN accelerator departments for the excellent performance of the LHC and thank the technical and administrative staffs at CERN and at other CMS institutes for their contributions to the success of the CMS effort. In addition, we gratefully acknowledge the computing centers and personnel of the Worldwide LHC Computing Grid and other centers for delivering so effectively the computing infrastructure essential to our analyses. Finally, we acknowledge the enduring support for the construction and operation of the LHC, the CMS detector, and the supporting computing infrastructure provided by the following funding agencies: BMBWF and FWF (Austria); FNRS and FWO (Belgium); CNPq, CAPES, FAPERJ, FAPERGS, and FAPESP (Brazil); MES and BNSF (Bulgaria); CERN; CAS, MoST, and NSFC (China); MINCIENCIAS (Colombia); MSES and CSF (Croatia); RIF (Cyprus); SENESCYT (Ecuador); MoER, ERC PUT and ERDF (Estonia); Academy of Finland, MEC, and HIP (Finland); CEA and CNRS/IN2P3 (France); BMBF, DFG, and HGF (Germany); GSRI (Greece); NKFIA (Hungary); DAE and DST (India); IPM (Iran); SFI (Ireland); INFN (Italy); MSIP and NRF (Republic of Korea); MES (Latvia); LAS (Lithuania); MOE and UM (Malaysia); BUAP, CINVESTAV, CONACYT, LNS, SEP, and UASLP-FAI (Mexico); MOS (Montenegro); MBIE (New Zealand); PAEC (Pakistan); MSHE and NSC (Poland); FCT (Portugal); JINR (Dubna); MON, RosAtom, RAS, RFBR, and NRC KI (Russia); MESTD (Serbia); MCIN/AEI and PCTI (Spain); MOSTR (Sri Lanka); Swiss Funding Agencies (Switzerland); MST (Taipei); ThEPCenter, IPST, STAR, and NSTDA (Thailand); TUBITAK and TAEK (Turkey); NASU (Ukraine); STFC (United Kingdom); DOE and NSF (USA).
  
  \hyphenation{Rachada-pisek} Individuals have received support from the Marie-Curie program and the European Research Council and Horizon 2020 Grant, contract Nos.\ 675440, 724704, 752730, 758316, 765710, 824093, 884104, and COST Action CA16108 (European Union); the Leventis Foundation; the Alfred P.\ Sloan Foundation; the Alexander von Humboldt Foundation; the Belgian Federal Science Policy Office; the Fonds pour la Formation \`a la Recherche dans l'Industrie et dans l'Agriculture (FRIA-Belgium); the Agentschap voor Innovatie door Wetenschap en Technologie (IWT-Belgium); the F.R.S.-FNRS and FWO (Belgium) under the ``Excellence of Science -- EOS" -- be.h project n.\ 30820817; the Beijing Municipal Science \& Technology Commission, No. Z191100007219010; the Ministry of Education, Youth and Sports (MEYS) of the Czech Republic; the Deutsche Forschungsgemeinschaft (DFG), under Germany's Excellence Strategy -- EXC 2121 ``Quantum Universe" -- 390833306, and under project number 400140256 - GRK2497; the Lend\"ulet (``Momentum") Program and the J\'anos Bolyai Research Scholarship of the Hungarian Academy of Sciences, the New National Excellence Program \'UNKP, the NKFIA research grants 123842, 123959, 124845, 124850, 125105, 128713, 128786, and 129058 (Hungary); the Council of Science and Industrial Research, India; the Latvian Council of Science; the Ministry of Science and Higher Education and the National Science Center, contracts Opus 2014/15/B/ST2/03998 and 2015/19/B/ST2/02861 (Poland); the Funda\c{c}\~ao para a Ci\^encia e a Tecnologia, grant CEECIND/01334/2018 (Portugal); the National Priorities Research Program by Qatar National Research Fund; the Ministry of Science and Higher Education, projects no. 14.W03.31.0026 and no. FSWW-2020-0008, and the Russian Foundation for Basic Research, project No.19-42-703014 (Russia); MCIN/AEI/10.13039/501100011033, ERDF ``a way of making Europe", and the Programa Estatal de Fomento de la Investigaci{\'o}n Cient{\'i}fica y T{\'e}cnica de Excelencia Mar\'{\i}a de Maeztu, grant MDM-2017-0765 and Programa Severo Ochoa del Principado de Asturias (Spain); the Stavros Niarchos Foundation (Greece); the Rachadapisek Sompot Fund for Postdoctoral Fellowship, Chulalongkorn University and the Chulalongkorn Academic into Its 2nd Century Project Advancement Project (Thailand); the Kavli Foundation; the Nvidia Corporation; the SuperMicro Corporation; the Welch Foundation, contract C-1845; and the Weston Havens Foundation (USA).
\end{acknowledgments}

\bibliography{auto_generated}

\cleardoublepage \appendix\section{The CMS Collaboration \label{app:collab}}\begin{sloppypar}\hyphenpenalty=5000\widowpenalty=500\clubpenalty=5000\input{SMP-20-015-authorlist.tex}\end{sloppypar}
\end{document}

%% file: SMP-20-015-authorlist.tex
\cmsinstitute{Yerevan~Physics~Institute, Yerevan, Armenia}
A.~Tumasyan
\cmsinstitute{Institut~f\"{u}r~Hochenergiephysik, Vienna, Austria}
W.~Adam\cmsorcid{0000-0001-9099-4341}, J.W.~Andrejkovic, T.~Bergauer\cmsorcid{0000-0002-5786-0293}, S.~Chatterjee\cmsorcid{0000-0003-2660-0349}, K.~Damanakis, M.~Dragicevic\cmsorcid{0000-0003-1967-6783}, A.~Escalante~Del~Valle\cmsorcid{0000-0002-9702-6359}, R.~Fr\"{u}hwirth\cmsAuthorMark{1}, M.~Jeitler\cmsAuthorMark{1}\cmsorcid{0000-0002-5141-9560}, N.~Krammer, L.~Lechner\cmsorcid{0000-0002-3065-1141}, D.~Liko, I.~Mikulec, P.~Paulitsch, F.M.~Pitters, J.~Schieck\cmsAuthorMark{1}\cmsorcid{0000-0002-1058-8093}, R.~Sch\"{o}fbeck\cmsorcid{0000-0002-2332-8784}, D.~Schwarz, S.~Templ\cmsorcid{0000-0003-3137-5692}, W.~Waltenberger\cmsorcid{0000-0002-6215-7228}, C.-E.~Wulz\cmsAuthorMark{1}\cmsorcid{0000-0001-9226-5812}
% ...
\vskip\cmsinstskip
% ...
1:~Also at TU Wien, Wien, Austria\\
% ...
